\documentstyle [12pt,aasms4]{article}
\input{epsf}
\journalid{}{}
\articleid{}{}
\begin{document}
\def\lax    {\ifmmode{_<\atop^{\sim}}\else{${_<\atop^{\sim}}$}\fi}
\def\gax    {\ifmmode{_>\atop^{\sim}}\else{${_>\atop^{\sim}}$}\fi}
\def\gtorder{\mathrel{\raise.3ex\hbox{$>$}\mkern-14mu
             \lower0.6ex\hbox{$\sim$}}}
\def\ltorder{\mathrel{\raise.3ex\hbox{$<$}\mkern-14mu
             \lower0.6ex\hbox{$\sim$}}}
\title{Do the Spectra of Soft X-ray Transients Reveal Bulk Motion Inflow 
Phenomenon?}

\author{ Konstantin Borozdin\altaffilmark{1,2},
Mikhail Revnivtsev\altaffilmark{2},
Sergey Trudolyubov\altaffilmark{2},
Chris Shrader\altaffilmark{3,4}
and Lev Titarchuk\altaffilmark{3,5} }
\altaffiltext{1}{NIS-2, Los Alamos National Laboratory,
Los Alamos, NM 87545; kbor@nis.lanl.gov}
\altaffiltext{2}{High--Energy Astrophysics Department,
Space Research Institute of Russian Academy of Sciences,
Profsoyuznaya 84/32, Moscow, 117810, Russia}
\altaffiltext{3}{Laboratory for High--Energy Astrophysics,
NASA Goddard Space Flight Center, Greenbelt, MD 20771, USA;
shrader@grossc.gsfc.nasa.gov,titarchuk@lheavx.gsfc.nasa.gov}
\altaffiltext{4}{Universities Space Research Association,Lanham MD}
\altaffiltext{5}{George Mason University/Institute for
Computational Sciences and Informatics, Fairfax VA}

\rm
\vspace{0.1in}
\begin{abstract}
We present our analysis of the high-energy radiation  from  black hole
(BH) transients, using archival data obtained primarily with the Rossi
X-ray Timing Explorer (RXTE), and a comprehensive test of the  bulk
motion Comptonization (BMC) model for the high-soft state continuum. The
emergent spectra of over 30 separate measurements of GRO J1655-40, GRS
1915+105, GRS 1739-278, 4U 1630-47 XTE J1755-32, and EXO~1846-031 X-ray
sources are successfully fitted by the BMC model, which
has been derived from basic physical principles in previous work.
This in turn provides direct physical insight into the innermost
observable regions where matter impinging upon the event horizon
can effectively be directly viewed. The BMC model is characterized by
three parameters: the disk color temperature, a geometric factor related to
the illumination of the black hole (BH) site by the disk and a spectral
index related to the efficiency of the bulk motion upscattering.  
For the case of GRO J1655-40, where there are distance and mass 
determinations, a self consistency check of the  BMC model has been made, 
in particular, the  assumption regarding the dominance of gravitational 
forces over the pressure forces within the inner few Schwarzschild radii.
We have also examined the time behavior of these parameters which can provide 
information on  the source structure.

Using our inferred model parameters: color temperature, spectral index
and an absolute normalization we present new, independently derived,
constraints on the black hole mass, mass accretion rate and the distance
for the aforementioned sources.  Also notable is  the relationship
between the color temperature and flux, which for GRO~J1655-40 is entirely
distinct from a simple $T^{4}$ dependence, and strikingly consistent
with the disk model we have invoked - standard Shakura-Sunyaev's disk
with the modification to the electron  scattering.
This provides insight into the origin of the seed soft photons, and
allows us to impose an important estimation of the hardness parameter, $T_{h}$,
which is the ratio of the color temperature to the effective temperature - 
we find $T_{h}\simeq2.6$, higher than previous  estimates used in the 
literature.
\end{abstract}


\section{Introduction}

Soft X-ray transients are sources of particular interest given their
unique properties: bright sources of this group appear on X-ray sky once
or twice per year,  typically produce  the flux from several hundred
milliCrab to several Crab in maximum and fade back to undetectable level
over time scales of weeks or months. During the outburst they undergo
major changes in their X-ray luminosity and spectral shape. Typical
spectral shapes for these sources  are well-known (see Sunyaev et al.,
1994; Tanaka and Shibazaki, 1996).
There are several virtually irrefutable black hole candidates among
these sources, based on quiescent optical observations. The binary mass
functions, which represent a lower limit for the compact object mass,
derived from such observations exceed upper theoretical limits for
neutron star mass in some cases, so these objects are widely believed to
be black holes (see e.g.  McClintock, 1998, for recent review).

Over the last decade, observations clearly demonstrated the existence of
at least two qualitatively different spectral states of Galactic black
hole binaries (e.g. White et al 1984; Grebenev et al., 1991, 1992;
Sunyaev et al., 1994; for recent review see e.g. Tanaka, 1997). A hard
power-law-like spectrum was first identified for the prototypical black
hole candidate Cyg X-1 (Tananbaum et al., 1972; Sunyaev \& Truemper,
1979).
It has subsequently been observed in many other systems. The commonly
accepted mechanism  for the generation of this hard radiation is thermal
Comptonization  (Shapiro, Lightman \& Eardley, 1976; Sunyaev \&
Titarchuk, 1980,  hereafter ST80).  Applying the thermal Comptonization
model to the hard-state BH spectra suggests the presence of a hot
cloud surrounding the central object (BH) with  typical plasma
temperature around 60 keV and Thomson optical depth of order of 1. Such
a cloud is able to Comptonize soft photons which originate in the
central region.
\par
For a number of years this spectral shape was considered to be a  BH
signature.  Subsequently, however, similar spectra  have been detected
from several X-ray bursters (Mitsuda et al., 1989; Barret \& Grindlay,
1995), and recently, from some X-ray pulsars (Negueruela et al., 1998;
 Gilfanov et al., 1998).
The luminosities of these neutron star binaries are
typically by an order of magnitude less than the Eddington limit.
Thus it is clear that there are common spectral features in the BH and
NS binaries under certain conditions.  In contrast, the high-soft state
spectra appear to be distinct. Neutron star binaries exhibit a soft
thermal component characterized by color temperatures $\simeq1-2$ keV,
which is interpreted as radiation emanating from the neutron star
surface or an accretion disk and/or boundary layer. In the BH case we 
have a thermal component from accretion disk and often see an additional
hard power law component dominating at energies higher than 10--20 keV.
\par
There have been attempts to explain hard emission component of the
high-soft state within the framework of the  thermal Comptonization
model which successfully describes the hard state spectra (for a recent 
review of models see Liang, 1998). We believe, that this approach 
encounters serious problems and cannot provide a clear physical picture 
without the inclusion of some rather ad hoc components. 
If one interprets the spectra observed 
as the result of the thermal Comptonization of soft disk photons in a hot
corona situated close to the cold disk, then typical power-law indices of
the high-soft state spectra, around 1.5, and high-energy cutoffs above
$\sim 300$~keV require quite small optical depths, $\simeq 0.1$, 
(Titarchuk \& Lyubarskij 1995,  see Fig. 2 and Eqs. 14 and 16).
The crucial question for such an interpretation is, 
how can one account for the observed stability of the spectral indices 
seen in the high-soft state spectra given the required 
high temperature of coronal plasma, $kT \simeq 150-200$~keV.
Taking into account the high efficiency of the energy amplification 
in such a hot corona, any variation in the optical depth by a factor 
of 2-3, which is natural to expect in a number of different sources, 
will lead to a detectable change of the spectral index, e.g. 
from  1 to  2.  We would routinely detect large
sporadic variations in the spectral slope, which is not what is
in fact seen in observations.  Another natural question is:
{\it why should extragalactic black hole sources, for which
physical scales differ by orders of magnitude, keep the same 
invariable optical depth in their coronae?} 
These and other difficult problems appear in connection with
the interpretation of the high-soft state spectra 
by the models of the thermal Comptonization.

An alternative understanding has been developed in a
series of papers, where the importance of bulk motion inflow in the
formation of emergent spectra was taken into account (Chakrabarti \&
Titarchuk 1995;  Titarchuk, Mastichiadis \& Kylafis 1996, 1997;
Titarchuk \& Zannias 1998, hereafter TZ98). Tests applying the BMC model
to data  have thus far been encouraging (Shrader \& Titarchuk 1998,
herein ShT98).
\par
In this paper we present results of our spectral analysis of X-ray
transients in the high-soft states in terms of the bulk motion
Comptonization (BMC) model. Over 30 spectra are utilized making this as
comprehensive an observational test as is currently plausible. We
describe observational data used and methods of analysis in \S 2.  The
main features of BMC model are outlined in \S 3.  Results of our data
analysis are presented in \S 4 and  \S 5 is devoted to the physical
parameter estimation based on our BMC model fitting results.
The issue of the origin of the seed soft photons is considered in
\S 6. The geometry of illumination of the converging site is discussed
in \S 7.  Discussions and conclusions  are offered in \S 8.

\section{Observations}

We have analyzed publicly available RXTE data for five transients,
which represent ultra-high spectral state of black hole candidates:
GRS1915+105, GRO J1655-40, GRS1739-278, XTE J1755-324 and 4U1630-47.
A list of observations is presented by Table 1 and the results
of the spectral analysis are shown in Tables 2-3.  We have
analyzed 10 separate observations of GRS1739-278, 20 observations of GRO
J1655-40, 2 observations of XTE J1755-324, 6 observations of 4U
1630-47.  Our analysis showed that GRS1915+105 represents more
complicated case, because this source undergoes dramatic changes in
intensity and spectrum during single observational sessions. Spectra
accumulated for periods of several seconds demonstrate quite noticeable
difference between this source and other systems discussed here.  We
believe, that the origin of these differences is due to a much higher accretion
rate in GRS1915+105, which causes the screening of black hole site by an
optically thick plasma cloud.  The unique properties of GRS1915+105 are
a subject in and of themselves, and further discussion is beyond the
general scope of this paper.

For broad band coverage of RXTE transients spectra we used the data from
both main instruments onboard the satellite - the PCA and HEXTE
spectrometers. For the processing of PCA data, the standard RXTE ``FTOOLS" 
software was applied.
Background estimation was carried out using the appropriate models for 
the bright or fainter objects as described by Stark et al. 
(1998 in preparation).  The comparison
between the modeled background spectrum and the observed one during
off-source pointings showed count-rate uncertainties at the level of 2-5
percent. We used the estimated background,  taking into account the
variation of the background over the orbit. We used PCA data in the band
from 3 to 20 keV, where effective area  is large enough and background
estimation is reasonably accurate.
Furthermore, to avoid the potential problems with associated
with the detailed structure of the soft disk-blackbody component we
restricted ourselves to the 5--20 keV energy band. To account for the
additional uncertainties in PCA response  matrix, we added a systematic
error of few percent per bin prior to spectral fitting. A deadtime
correction was calculated according to Zhang et al., 1995.
For HEXTE data analysis we used the version of response matrix released
on April 3, 1997 and standard off-source observations of each cluster
of detectors to subtract the background.  The spectral band used
was 15-150 keV, and a deadtime correction was applied.
HEXTE data were included in our subsequent analysis only when they were
statistically significant. The HEXTE data have been re-normalized based
on the PCA. Finally, we note, that that there are known to
be small cross-calibration discrepancies between the HEXTE and PCA
instruments, but we do not believe that these are at a level which is likely to
effect our basic conclusions.

In addition to the RXTE sample, we obtained archival data
for covering of  the April 1985 outburst of the transient EXO1846-031
with EXOSAT  (Parmar et al. 1993).
These data were obtained with the GSPC detector, covering a
2-20 keV bandpass, with a nominal energy resolution of $E/\delta E
\simeq 5$.
Although we do not have as large number of spectra  as in several of the
objects observed with RXTE, the addition of this object  complements the
RXTE sample in several important ways. For one, it is  useful
cross-check to further test the model using data from a  separate
instrument. Furthermore, EXO 1846-031 has an outburst history  distinct
from that of the other sources in our sample -- it was detected  in the
EXOSAT era, and has not undergone subsequent activity like for  example
GRS 1915+105 which has essentially ``stayed on" following its  original
outburst -- it is thus in some sense more representative  of the broader
class of X-ray nova (e.g. Chen, Shrader \& Livio 1997).

In addition to the RXTE and EXOSAT data, we have examined the
higher-energy range covered by the Oriented Scintillator Spectrometer
Experiment (OSSE) on board Compton GRO for high-soft-state
transient episodes in two sources: GRO~J1655-40 and XTE~J1550-564.
These data were obtained from the public archives; there were no
concurrent soft-X-ray observations for the GRO~J1655-40 data in
question, and we do not currently have access to simultaneous Soft X-ray data 
for XTE~J1550-564.
\par
Our purpose in this case was to study the extent of the hard power law.
GRO~J1655-40-57 was observed with {\it OSSE} during {\it CGRO}
on a number of occasions in 1994 and 1996 (Grove et al.1998;
Tomsick et al. 1998). The Burst And Transient
Source Experiment (BATSE) monitoring data, covering the nominal
20-200 keV energy range were examined to gauge the relative hard-X-ray
intensity at the time of these observations. This led us to consider
the OSSE data from viewing periods 405.5 (MJD~49693-49699)
and 414.3 (MJD~49805-49811). In addition, we obtained the public OSSE
data for XTE~J1550-564 (Smith 1998).
In each case, the data utilized were obtained in standard ``chopping''
mode with about 450  energy channels per detector covering the  50 keV - 
10 MeV band with a  nominal energy resolution of $E/\Delta E \simeq 12$ at 
200 keV (for a technical description see Johnson et al. 1993).  
The standard $\sim2$-minute on- off-source
accumulations were analyzed using the {\it OSSE} analysis software
package,``IGORE'', incorporating the default calibration
and background model parameters. Data summation, and response matrix
generation were also performed using IGORE, and the resulting count
spectra and response matrices were then imported into XSPEC for
subsequent model fitting.

In our subsequent analysis, we are interpret the spectra thus obtained
within the framework of bulk motion Comptonization (BMC) model.
 Analytical approximations of the model
(Titarchuk, Mastichiadis, \& Kylafis 1996, 1997, hereafter TMK;
Titarchuk \& Zannias, 1998, hereafter TZ98), the validity of which
has been confirmed through Monte Carlo simulations by Laurent \& Titarchuk 1998
 (hereafter LT98), have been imported into the ``XSPEC''software package with 
which our spectral deconvolutions have been performed.

\section{Bulk-Motion Spectral Model}

For our interpretation we assume the following overall geometry of  the
system.  The plasma taken from optical companion is accreting as a sum
of two components - one, which is Keplerian, forming an accretion disk
and another, which is non-Keplerian, leading to a quasi-spherical
accretion component that may eventually be stopped by a centrifugal barrier
 - or it may just smoothly  proceed towards a black hole releasing energy due
to Coloumb collisions, geometric compressions (Chakrabarti \& Titarchuk,
1995, hereafter CT95)  and viscous dissipation (Narayan \& Yi 1994).
Neither of these components can exist in the closes-in environment of
the black hole,  where nothing can stop the matter in its relativistic
movement towards the event horizon. The regions involved in the X-ray
emission are in this case 1) the internal part of the ``classical''
accretion disk; 2) a high-temperature region, the Compton cloud; and 3)
internal bulk-motion Comptonization (BMC) region. The
presence of the disk,  itself is a sign  that the radiation pressure in 
the system is much less the
Eddington value - by factor of $H/R$ (where $H$ and $R$ are a disk
half-thickness and radius respectively). In fact, the hydrostatic
equilibrium in the disk can be sustained only  if the local luminosity
in the disk is an $\sim H/R~L_{edd}$.   Thus, the radiation cannot hold
the matter  situated outside the disk from entering in an almost free
fall manner onto the  black hole horizon. The accretion disk is believed
to emit as a multicolor blackbody [see Shakura \& Sunyaev (1973),
hereafter SS73]. A high temperature region, which we refer to as the
Compton cloud, is usually associated  with the generation of Thermal 
Comptonized emission in the hard-low state. The Compton cloud however, will be
substantially cooled by powerful flux of low energy photons from the disk, when
the accretion rate increases and the system approaches its high-soft
spectral state.  The bulk motion region then becomes a principal source
of hard X-ray photons.
\par
It has been shown elsewhere (TMK97, TZ98, LT98) that two effects,
the bulk motion upscattering and  the
Compton (recoil) downscattering, compete forming the hard tail of  the
spectrum as an extended power law. The soft part of the spectrum  comes
from the disk photons seen directly and a subset of those photons, which
escape from the BMC atmosphere after undergoing a few scatterings  but
without any significant energy change. It has also been shown that, even
neglecting special and general relativistic effects,
one is able to reproduce the main features of the full  relativistic
formalism: the overall spectral energy distribution and  the dependence
of the high-energy power law on mass accretion rate (TMK97, Monte-Carlo
calculations in LT98).

It is worth noting that BMC model predicts a specific feature - spectral
turnover at the energies near the electron rest mass $\sim 500$ keV.
The exact energy of the turnover is  determined by several effects
which can shift it in either direction.
The turnover energy is a function of the plasma temperature
which becomes increasingly sensitive for temperatures higher than 10 keV
(LT98).
There is also a possibility that  the electron
distribution in the converging inflow can deviate  from a Maxwellian,
flattening at high velocities, since there is insufficient time for it
to thermalize (Ellison, Baring \& Jones, 1996). The hard photon
tail  may then extend to higher energies.
On the other hand, due to gravitational
redshift and photon bending effects, the injected soft photons
are not Comptonized very effectively.
The hard tail of the BMC spectrum is quite steep
-- the energy spectral index is higher than 1 -- and there is a
notorious dearth of (net) photons in the low-energy gamma-ray band.
Since the extent of the high-energy power law comprises a specific diagnostic
of the BMC model, we examine what constraints current observations
can impose in Section 5.

Any high-energy spectral feature related to a relativistic
electron distribution, for example that which is presumably
responsible for the appearance of jets,  is not uniquely
constrained to this energy band, i.e. not tied to $m_ec^2$.
If the power-law component, or some other spectral feature is
found to extend to energies $\sim 1$ MeV, one should consider
alternative origins for this emission, such as relativistic jets. 
These are far less
restricted in terms of possible Lorentz factors and thus potentially
able to generate photons with extremely high energies.

Unfortunately, the data obtained so far for energies near
1 MeV are severely background limited (Grove et al. 1998)
and cannot address this issue
in a definitive manner. We hope that future gamma-ray missions
(e.g. INTEGRAL) will be able to resolve such issues and
help to delineate between the models. A more detailed discussion
of BH spectra in gamma-ray band is well outside of the scope of this
paper.
\par
We have assumed that the external illumination of the converging (BMC)
flow is due to low-energy thermal radiation from the accretion
disk characterized by a temperature $T_{col}$.  This assumption is model
independent.  We show below (\S 5) that the multicolor disk  model is
reasonably well described by this shape in the energy band  where the
photon energies are always higher than the characteristic energies of
the soft injected photons. This assumption is valid for the RXTE energy
band (which starts from 3 keV) because the soft photon color
temperature is always $\sim1$ keV  or less (some important exemptions
will be discussed in the next section).
\par
Furthermore, we have assumed that this illuminating radiation impinges
on the BMC atmosphere in a manner characterized
by a certain geometry, which we have paramaterized in terms
of a geometric parameter $f$. This parameter is
really the first expansion coefficient of the spatial source photon
distribution over the set of the eigenfunctions of the BMC formulation
(TMK97, Eq. 30). It is easy to prove (using the property of
the upscattering Green's function to conserve the number of scattering
photons) that {\it $f$ is exactly the ratio of the number of photons
multiply scattered in the converging inflow to the number of photons in
thermal component}. As it is clearly demonstrated in TMK97, TZ98 and
LT98 the spectral index  is independent of the illumination parameter,
$f$.
\par
In \S 7  we discuss the possible scenarios  of the illumination of
the converging inflow region by the soft photons of the disk and some
useful formulae for the factor $f$ are derived.

In the soft state when the accretion rate is higher, photons from
the Keplerian disk cool the ambient Compton cloud through thermal
Comptonization and free-free emission (CT95). The cooler converging
inflow, as it rushes towards the black hole,  scatters soft-photons
within the radius, $r\sim \dot m r_s$ --  some of the photons  then
undergo outward radiative diffusion. Here $\dot m=\dot M/\dot M_E$,
$r_s$ is Schwarzschild radius, $\dot M$ is the net accretion rate
(including accretion from the disk plus any halo or other non-keplerian
component), $\dot M_E \equiv L_E/c^2=4\pi GMm_p/ \sigma_Tc~$ is the
Eddington accretion rate, $M$ is the mass of the central object, $m_p$
is the proton mass and $G$ is the gravitational constant. Momentum is
transferred to the soft-photons producing a power-law component
extending to energies comparable to the kinetic energy of electrons in
the converging inflow, i.e. of order $m_ec^2 = 511$ keV.
As we mentioned above the luminosity of the upscattered component,  the
hard power law, can be very small compared to the Eddington luminosity.
\par
On the other hand {\it in the hard state, the hot Compton cloud
obscures the BMC region and no upscattered photons are seen}.
\par
The BMC spectral model can be described as the sum of a thermal (disk)
component and the convolution of some fraction of this component
$g(E_0)$ with the upscattering Green's function $I(E,E_0)$ (TMK97, Eq.
30). The Green's function has the form of a broken power-law with
spectral indices $\alpha$ and $\alpha+\zeta$ for high  $E\geq E_0$ and
low $E\leq E_0$ energy parts respectively,
\begin{equation} F_{\nu} (E)=\int_0^{\infty}I(E,E_0)g(E_0)dE_0.
\end{equation}
\par
\noindent
The above convolution is insensitive to the value of the Green's
function spectral index $\alpha+\zeta$, which is always much greater
than one. The validity of this formalism has recently been confirmed
through extensive Monte Carlo simulations (LT98).
Furthermore, TZ98, using the full relativistic formalism, have
demonstrated that for a Schwarzschild black hole, the
extended power law is formed in the bulk motion inflow,  the production
of the hard photons as a function of radius peaks at $\simeq2 R_s$
($R_s$ is Schwarzschild radius) and that  most of the hard photons
observed are produced within $3 R_s$.
Furthermore, the same statement is valid for the case of a  rotating
(Kerr) black hole, although a higher mass accretion rate is required to
provide the same efficiency for the soft photon upscattering.

For low plasma temperatures, one obtains a power-law energy
index $\simeq1.8$,  which we note is the canonical value for the Narrow
line Seyfert Galactic Nuclei (Brandt, Mathur \& Elvis 1997).
Furthermore Laurent \& Titarchuk 1998, using Monte-Carlo
calculations, have shown that the hard-soft transition can be achieved
through the plasma temperature  regulation in the bulk motion
inflow from a few keV up to 50 keV. They have also shown that for the
low temperature limit ($<10$ keV) the hard  tail is formed,
but when the temperatures approach 50 keV, the spectrum is almost
indistinguishable from the thermal Comptonization case.

We note that the processes of absorption and emission, as free-free or
synchrotron radiation, can be neglected provided the plasma temperature
of the bulk flow is $\sim1$ keV or greater for characteristic number
densities $\sim10^{18}$ cm$^{-3}$ and for magnetic field strengths
in the proximity of the black hole $\sim10^4-10^5$ Gauss or less.

\section{Spectra}

Examples of our spectral fitting  are presented in Figure 1.
Application of the BMC model demonstrates good agreement with the data,
even though the spectrum of accretion disk is approximated
by a single-temperature black-body.
For spectra extending below several keV,  replacement of
the single black-body by a multicolor disk component within the
framework of BMC model is expected to be necessary, but
given the limited soft sensitivity of RXTE, the single black body
approximation is adequate for most spectra we have analyzed.

\centerline{Editor: Place Figure 1 here}

As we have noted, significant deviations from a single-temperature black
body shape are expected in the emergent spectra due to electron
scattering effects. This emergent ``diluted''
blackbody spectral form has been observed in GRS1915+105, GRO J1655-40
and 4U1630-47. A convolution of a black body function with Green's
function is needed to satisfactorily describe these spectra. One
example, 4U1630-47, is illustrated in Fig.1.  We believe, that
in this case the soft photons site is screened by a plasma
cloud, which dilutes incident spectrum with a net effect of shifting it
to higher energies.  Applying this
assumption we are able to determine the temperatures characterizing the
incident photon distribution and the cloud from the inferred parameters
of our fit. Typical cloud temperatures are $\sim5$~keV have thus 
far been obtained.  We plan  to study such
spectra in more systematic manner and present these results elsewhere.
Here we have concentrated on the more simple case, where the black body
approximation is able to adequately represent the data.

In some cases these spectra can be satisfactorily fitted by the sum of
two components: a multicolor disk component and a power law. Our
approach has however, two important advantages. First, if one
uses a simple sum of two such components, it is not known, how far the
power law extends towards low energies. This lack of knowledge can
potentially cause incorrect physical information to be inferred the
parameters of the fit; notably the disk effective temperature. Thus,
estimations of the total luminosity and energy inputs from both
components can be inaccurate.  Second, the BMC model parameters are physically
meaningful, and from them one can derive important constraints on the
geometry of the system, the central-object mass, the mass accretion rate
and the distance to the source (see \S 5).

\centerline{Editor: Place Figure 2 here}

The RXTE coverage of GROJ1655-40 provides a good sample of
high-state spectra. The variation of our derived parameters with time is
presented in Fig.2.   The correlation between the color temperature
and X-ray flux is evident (see also Fig. 7 ),  which is quite natural,
since most of the observed photons  come from the accretion disk.
However, this correlation cannot be approximated by single 
$F \sim T^4$  function as one would expect in the case of constant 
effective area (a similar conclusion for GRS1739-278 is demonstrated 
in Borozdin et al., 1998).
These results demonstrate that the effective radius of the accretion
disk is variable and depends on variations in color temperature 
(\S 6 addresses this issue in detail).
This became evident for the last five observations of this series where
we see the beginning of the transition to the low-hard
state, accompanied by a drastic decrease in color temperature. The
soft component is no longer detected during following observations
[see \cite{trud98}, \cite{sob98}].

Another important result is the lack of an overall correlation between
the disk color temperature and the power-law  index. We see such a
correlation, for example for the observation of March 24, 1997, when
the drop in temperature corresponds to a steepening of the power-law.
This is not the case, however, for other periods when the
power-law became steep without any noticeable change of the temperature,
namely, for a series of sessions between
Apr 24 and May 20, and on June 26.  We believe, that during these
periods of time there might be some {\it outflow} from the black hole
site, which reduces the amount of matter falling into the BMC region.
Or there is another possibility that  the increase of the spectral index
(or softening of the spectrum) during the soft-high
state can be also related  with the downscattering degradation of the
hard photons (produced in the very core of the bulk inflow region) 
when these photons escape towards an observer passing through the relatively 
cold optically thick material of the expanded disk during the flares.

To confirm this interpretation we have examined some radio
measurements,which would support the presence of outflowing plasma during those
periods.

We have studied publicly available data for GRO J1655-40 obtained
with the Green Bank  Interferometer at 2.25 GHz frequency (The Green
Bank Interferometer is a facility of the National  Science Foundation
operated by the NRAO in support of  NASA High Energy Astrophysics
programs).  These data, which are integrated for several days per point,
are shown at the lower panel of Fig.2.  The data show no significant
increase of activity during the entire period of RXTE observations,
although it appears that the highest fluxes were detected concurrently
with an increase in the power law index.  These  results support our
interpretation, but unfortunately are not sensitive enough to make any
strong conclusion.  Observations with more sensitive instruments are
needed to confirm or disprove this correlation.

For GRS1739-278 the data sample is not so evenly spaced as for GRO
J1655-40. The first observation, March 31, 1996,  was carried out when
the system was in an ultra-high spectral state with a spectrum formed by
the two strong components: the accretion disk component and extended
power-law seen in energies from 10~keV and up to 100~keV.
In May 1996, a series of observations was carried out, where the
power-law component was much weaker and the source was detected by only
by the PCA experiment.  The BMC model was able to satisfactorily describe the
data in both these cases. Derived parameters, as was the case for GRO
J1655-40, show a correlation between disk temperature and flux,
 which is evidently related to variations in accretion rate. 
Other dependencies are less evident, because of the weak significance of 
the power-law component in most cases examined.

As noted in Section 3, an important diagnostic of the BMC model
is the behavior of the high-energy power law; specifically, a
break at energies $\sim500$~keV is predicted. As part of our
comprehensive study of the BMC model and its application to
current observational data, we have explored what constraints can
or cannot be imposed.

\centerline{Editor: Place Figure 3 here}

In Figure 3 we have plotted the
background-subtracted count spectra resulting from the GRO~J1655-40 (a)
and XTE~J1550-564 (b) described in Section 2. We have plotted the
count-rates per detector-channel superposed on curve of zero intensity.
Only channels above 45, corresponding to approximately 300 keV are
shown. The data are consistent we believe with any significant lack 
of flux above 500~keV in each case.
We note that others, (Grove et al. 1998; Tomsick et al. 1998) have found
that the power-law {\it fit} that they obtain for GRO~J1655-40 extends
to energies as high as $\sim800$ keV, and that in fact this pure
power-law fit may lead to an improved chi-square relative to a
a model invoking a high-energy cut-off. Indeed, our own
fits to the data led to similar power-law indices and a positive
residual in the {\it model} between 500 and 600~keV, but in no
case is the statistical significance greater than 2 standard deviations.

We consider the issue to be unresolved, and cite the simple lack of
net-detector counts above 500~keV as the primary basis for our claim.
Future instrumentation, such as the Integral Spectrometer (SPI) may 
be able to unambiguously address this issue.

\section{Estimation of Physical Parameters of Black Hole Systems}

In this section we demonstrate the principal abilities of the BMC model
to estimate the values of physical parameters of the system. We also
show what kind of uncertainties are present in these determinations and
some ways they might be overcome.

Using derived parameters, the color temperature  and an
absolute normalization, one can infer the effective area into a
soft-emission region  if the distance to the source and the hardening
factor, $T_h$ (ratio of the color temperature to the effective
temperature)  are known.   This involves  translating  the observable
surface area to the black hole mass in terms of a specific disk model.
We have selected the SS73-model with a modification to treat electron
scattering  (e.g. Titarchuk 1994a) as our basic disk model.  The spectrum
of such a disk can be represented by the integral of a
diluted blackbody function characterized by  the dilution coefficient
$T_h^{-4}$ over  the entire disk.  The flux density $g_{\nu}$, measured
in units of erg~sec$^{-1}$~ster$^{-1}$cm$^{-2}$, is given by:
\begin{equation}
g_{\nu}(E)= C_{N}\cdot
2\int_{r_{in}}^{\infty}{{E^3(1.6\cdot10^{-9})}\over{\exp[E/T_{max}f(r)]-1}}rdr
\end{equation}

Here the blackbody disk color temperature distribution
$T_{max}\cdot f(r)$ is a function  of the radius $r=R/R_s$,
expressed in the units of Schwarschild radius $R_s=2GM/c^2$:
\begin{equation}
f(r)={{[1-(r_{in}/r)^{1/2}]^{1/4}(1.36r_{in})^{3/4}}\over{r^{3/4}(1/7)^{1/4}}},
\end{equation}
It is normalized to the maximum value $T_{max}$ at
$r=(7/6)^2r_{in}=1.36r_{in}$.
The normalization factor of the equation (2) measured in units
phot~sec$^{-1}$~ster$^{-1}$~ cm$^2$ is:
\begin{equation}
C_{N}={{0.91m^2\cos{i} T_h^{-4}}\over{d^2}},
\end{equation}
where $\cos i$ is the cosine of the inclination angle
(the angle between  the line of sight and the normal to the disk), the
black hole mass is $m=M/M_{\odot}$, and the distance to the source d is
in kpc units.

It is possible to replace the multicolor disk spectrum by the
diluted blackbody spectrum with the color temperature $T_{col}$  using
the theorem of the mean value (e.g. Korn \& Korn 1961), i.e.
\begin{equation}
 g_{\nu}(E)~=~A_{N}{{E^3(1.6\cdot10^{-9})}\over{\exp[E/T_{col}]-1}},
\end{equation}
\par
\noindent
where
\begin{equation}
A_{N}= C_{N}r_{eff}^2.
\end{equation}

It is worth noting that {\it in principle} $r_{eff}$ is a function of
the photon energy $E$. But if the energy band covered is such that ratio
of photon energy to  the blackbody temperature is greater than 2, the
effective radius is almost constant. In this case,  the blackbody
spectral shape with color temperature $T_{col}$ provides a reasonable
approximation to the multicolor disk spectrum (see Fig. 4).

\centerline{Editor: Place Figure 4 here}

In fact, the shape of the multicolor disk
spectrum is also  characterized by only one parameter, the  maximal
temperature $T_{max}$ (if the inner  edge of the accretion disk is fixed
at 3 Schwarzshild radii):
\begin{equation}
T_{max}={{3\cdot7^{-1/4}}\over{[(7/6)^2r_{in}]^{3/4}}}
\left({{\dot m_{disk}}\over{m}}\right)^{1/4}T_h~~{\rm keV}.
\end{equation}

>From the fitting of data with the BMC model one can obtain a color
temperature $T_{col}$ and an absolute normalization $ A_{N}$. Then,
for a given $T_{col}$, equations (2) and (5) give $r_{eff}$
and $T_{max}$.

\centerline{Editor: Place Figure 5 here}

\centerline{Editor: Place Figure 6 here}

In Figures 5 and 6 the dependencies of  $r_{eff}$,
$T_{max}$ and $T_{max}/T_{col}$ are presented  as functions of
$T_{col}$. These relations are calculated by using the least  squares method,
comparing the deviation between the multicolor disk and single
temperature blackbody spectra (equations 2 and 5 respectively).
It is worth  noting that higher color temperatures require higher
effective radii, i.e. at higher temperatures an effective area of
the disk is more outwardly extended. The ratio $T_{max}/T_{col}$ 
is in the range 1.15 and 1.3 when the color temperature changes 
from 0.7 keV to 1.1 keV.
Once  the best-fit $A_{N}$ and $r_{eff}$ are determined
one can use these values for the BH mass and the distance 
determinations(see Eq. 6).
\par
On the other hand  the mass accretion rate $\dot m_{disk}$ is related to
the BH mass, $m$ and $T_{max}$ through the relation:
\begin{equation}
\dot m_{disk} = m{{(T_{max})^4[(7/6)^2r_{in}]^3}\over{(1/7)3^4T_h^4}}.
\end{equation}
This allows one to find the mass accretion rate and ultimately
luminosity of the disk (soft) component.

\subsection{Constraints on the mass, mass accretion rate and the
distance of BH sources}

\subsubsection{Hardening factor determination for GRO J1655-40}

To check the self-consistency of this procedure we first make the
appropriate estimates for GRO J1655-40, since we know its
distance $d\approx3.2\pm 0.2$ kpc (Hjellming \& Rupen 1995), mass
$m\approx7$ and inclination angle $70^{0}$ (Orosz \& Bailyn 1997) to
a relatively high-degree of precision.

Equations (2)-(5) (see also Fig. 5) give us an effective radius
of the emission region in the disk which is related to
the color temperature $T_{col}$.
The best-fit parameters, $T_{col}$ and $A_N$,
along with the known mass, distance and inclination angle allow
us to determine the hardening factor $T_h$ from equation 6.

We then obtain a mass accretion rate $\dot m_{disk}$ from equation (8)
using the values for $m$, $T_{max}$ and $T_h$.
Then the luminosity of the soft component is determined as
\begin{equation}
L_s=1.4\cdot 10^{34} (\pi^5/15) m^2 r_{eff}^2
\left({{T_{col}({\rm keV})}\over{T_h}}\right)^4 ~~~~~~ {\rm erg~s^{-1}.}
\end{equation}

We have applied our inferred spectral parameters, using Eq.(6) and
a least-squares procedure, to obtain the mean value, standard deviation
and probability density distribution $\Psi(T_{col})$ for $T_h$.
It is easy to show using the least-squares method that
\begin{equation}
\Psi=r_{eff}^2(T_{col})/\Sigma r_{eff}^2(T_{col}^{int}),
\end{equation}
where the sum is implemented over all the best-fit color
temperature values $T_{col}^{int}$. For each set of best-fit parameters
$T_h^4=0.91\cdot (m/d)^2\cos{i}(r_{eff}^2/A_N)$ (see Eq.6).
Averaging this quantity over the probability distribution $\Psi$
one can find  the mean value and  standard deviation of $T_h$.

For the sample of 20 spectra of GRO J1655-40 we obtain
$T_h=2.57\pm 0.1$. From this result we can conclude that
the hardening factor for this sample is almost constant and
independent of color temperature.
This value of the hardening factor is consistent with
theoretical calculations by Shimura \& Takahara (1995),
provided the $\alpha$-parameter is $\gtorder$0.1,
i.e if the Reynolds number $\alpha^{-1}$ is $\ltorder$10.
In fact, Titarchuk, Lapidus \& Muslimov (1998) found that these
values of the $\alpha$-parameter are typical for the binary systems
found to exhibit kHz QPOs.

\subsubsection{Distance and BH mass  determination}

Unfortunately, for sources we lack the precise measurements
of the binary system that we are afforded with GRO J1655-40.
However, we can derive the distance-mass relationship for any BH system,
using parameters inferred from the BMC model, and assuming that
the hardening factor obtained above for GRO J1655-40 represents
a canonical value. We argue that the general use of that value is
justified,
because it was obtained over relatively wide dynamical range of color
temperatures.

For a given set of inferred parameters $A_N$, $T_{col}$ (and hence
$r_{eff}$), the distance-mass ratio is determined from equation (6) as:
\begin{equation}
\left({{d}\over{m}}\right)^2
={{0.91 T_h^{-4}r_{eff}^2}\over{A_N}}\cos{i}.
\end{equation}
Using the same averaging technique, we applied earlier
to determine of the mean value and standard deviation
of the hardening factor, we can find the mean value and standard
deviation of the distance-mass ratio.
With the assumptions that $T_h=2.6$ and $\cos{i}=0.5$, we find that
$d/m= 0.72\pm 0.09$,  $0.89\pm 0.06$, $0.95\pm 0.06$ for 4U 1630-47,
GRS 1739-278, and  XTE J1755-324 respectively.

The spectral indices
and the results of Monte Carlo simulations for the spectral index-mass
accretion rate relationship (see LT98, table 2) afford us the
possibility to elaborate the constraints on the mass and distance
separately. With the assumption
that the electron temperature of the converging inflow is low
($\ltorder$ 10 keV) we get an inequality for the mass accretion
rate in the converging inflow $\dot m$ which is a sum
of the disk mass accretion rate $\dot m_{disk}$ and mass accretion
rate of the halo (advection dominated) component. Namely,
$\dot m=\dot m_{disk} +\dot m_{halo}\gtorder 2$ and
$\dot m \gtorder 1$, if the spectral indices satisfy $\alpha\leq 1.7$
and $1.7<\alpha\leq 2.2$ respectively. Equation (8) then allows us
to place constraints on the  black hole mass with certain assumptions
regarding the hardening factor $T_h$ and the innermost disk radius
$r_{in}$.
For the former, we can use the value 2.6 obtained for GRO J1655-40 (see
above) and for the latter one we can use the radius of the last stable
orbit in the disk (r=3). Then the constraints on the distance can be
obtained using equation (10) with the absolute normalization
$A_N$ and the effective radius $r_{eff}$.

For mass-distance estimations we rewrite equations (8) and (10) as:

\begin{equation}
 m = \dot m_{disk}{{(1/7)3^4T_h^4}\over{(T_{max})^4[(7/6)^2r_{in}]^3}} =
m^{\star} \left({{T_h}\over{2.6}}\right)^4,
\end{equation}
and
\begin{equation}
 d = d^{\star} \left({{T_h}\over{2.6}}\right)^2
\left({{\cos {i}}\over{0.5}}\right)^{1/2}.
\end{equation}

The luminosity of the soft component (see Eq. 9) expressed in  units of
erg/s is determined by
\begin{equation}
L_s=1.4\cdot 10^{34} (\pi^5/15) m^2 r_{eff}^2
\left({{T_{col}({\rm keV})}\over{T_h}}\right)^4 =
10^{37}\cdot L_{37}^{\star}\left({{T_h}\over{2.6}}\right)^4.
\end{equation}

Here $m^{\star}$, $d^{\star}$ and $L^{\star}_{37}$ are functions of
$T_{col}$, $r_{eff}$ and $\dot m_{disk}$ and
represent the estimations or constraints we get for mass, distance and
soft component luminosity taking into account all the discussion above.

In table 4  we present the set of  parameters, obtained from our
spectral fitting, that we used for our parameter estimations
and the constraints inferred from equations (12)-(14) for
each source in our sample other than GRO~J1655-40.
\par
Fit parameters for GRS 1915+105 are taken from ShT98
 and we derive that the distance-to-mass ratio $d/m$ is 0.67 with an
assumption $T_h=2.6$
Using millimeter observations of the relativistic ejections in GRS
1915+105 Mirabel \& Rodrigues 1994  (see also Chaty et al. 1996) came to the
conclusion  that this source is at the distance $12.5\pm 1.5$
kpc from the Sun, behind the core of a molecular cloud at $9.4\pm 0.2$
kpc.
Applied this distance estimate we find that {\it the BH mass in GRS
1915+105 system is $18.6\pm 2.2$ solar masses}. 
We will verify these mass-distance constraints in our future
more extended data analysis of RXTE observations of GRS 1915+105.
 \par
XTE J1755-32 represents a special case, because its spectral index
$\alpha=1.0$ is quite low. This means that the converging inflow,
with a total mass accretion rate presumably higher than 2, is
not cooled effectively by the disk photons and that
$\dot m_{halo}$ may be comparable to, or even exceed the disk mass
accretion rate $\dot m_{disk}$. To estimate $\dot m_{halo}$ in this case
we have used the distance arguments presented by Trudolyubov et al 1998.

They have demonstrated that bright X-ray
transients tend to group around Galactic Center, where mass
density is maximal.  In the case of XTE J1755-32 its position on the
sky allows us to assume, that this is one of sources, which are not
too far away from the Center of Galaxy, and the distance is somewhere
between 6 and 12 kpc with high probability.
Thus with an assumption $d\sim 8$ kpc one can get from
the above estimates of the distance-mass ratio that a black hole mass
$m\sim 8.4$ and then using Eq. (12) one can get  $\dot m_{disk}\sim
0.5$.

\section{On the origin of the seed soft photons component}

As we have shown, the spectral energy distribution
of Compton scattered photons from the converging flow is very close
to what is observed.  But it is still necessary to understand the origin 
of the soft  component. If these soft photons come from the accretion disk, 
then what is the specific nature of the disk?  There are a lot of disk
models presented in the literature. Currently, among the most widely
discussed are those invoking an advection-dominated accretion disk
(see e.g. Narayan \& Yi 1994,  Esin, McClintock, \&  Narayan  1997,
Narayn, Barret, \& McClintock 1997). Other recent models are based
on the presence of two flows simultaneously proceeding
towards the central objects in the non-Keplerian and Keplerian
manners (e.g.  Chakrabarti 1990, 1996, Molteni, Sponholz, \& Chakrabarti
1996, CT95).
\par
In the previous sections, our analysis of the BH parameter
constraints were based on the standard SS73  disk model modified to
incorporate electron scattering effects. A question of major importance
is whether or not we can derive  any argument favoring one particular 
disk over another model from available data.

Here we attempt to  demonstrate how one can establish what is
the appropriate disk model from the apparent  color temperature of the
soft photons and the observable flux. Fortunately, GRO J1655-40 provides
a relatively large dynamical range of the color temperatures and fluxes.

\centerline{Editor: Place Figure 7 here}

In Figure 7 we present the observational relation between these two
quantities.  We remind the reader that the temperature of the black body
with a fixed surface area depends on the energy flux as the fourth
root.  It is evident from the figure that the data deviate
significantly from this dependence (dashed line). As a comparison, we
calculated the dependence of the color temperature on
the energy flux for the disk model detailed in the previous sections.
For a given color temperature $T_{col}$ we have found an effective
radius of the emission disk region $r_{eff}$, and then derived an
energy flux in the PCA band, integrating the blackbody spectral
distribution (see Eq. 5).

The results of these calculations are represented by the solid line and
two dotted lines, which represent the error corridor
(about 20\%) expected from a variety of possible uncertainties.
The theoretical dependence of the color temperature on flux was
obtained assuming the constancy of the  hardening factor in the given
range of the color temperatures.
Almost all of the points for GRO J1655-40 follow the theoretical curve
through one decade  in energy flux. This evident agreement of the
theory with the observations give us a solid basis for use of the
model which in turn allows us to constrain the BH system parameters.
The fact that the dependence of the flux on the hardening factor,
$T_h$ is very strong (it is proportional to $T_h^{-4}$),
yet the theoretical curve for a constant $T_h$ describes the observational 
points quite well, gives us confidence to fix the value of the
hardening factor in treating other sources within our sample.

In Figure 7 we present our flux and temperature determinations
for GRS 1739-278 source (hollow circles). Flux values
were scaled to approximate  GRO J1655-40 values.
GRS 1739-278 data are also consistent with
multicolor disk model and inconsistent with single-temperature
blackbody model (dashed-line).
Unfortunately, for the other sources we do not have comparable
statistics as in the case of GRO J1655-40. For those cases, arguments
characterizing the disk must be considered in terms of the similarity of their 
observational  appearances (e.g. color temperatures, luminosities {\it etc}). 
Further investigations are required to determine  the dependencies of
color temperature on flux for the more general case.

\section{Soft photons illumination of the converging inflow region}

Our disk model predicts that the emission surface
area is related to the color temperature of the soft spectral component
(Fig. 5); the emission region expands outward with increasing color
temperature. Thus one can expect a decrease in the illumination
parameter, $f$, with increasing color temperature, provided the
converging inflow (CI) remains situated inside 3 Schwarzschild radii. I
n this case the CI illumination is becoming less effective as the outer edge 
of the disk, characterized by $r_{eff}$, propagates  to larger radii.
If the converging inflow atmosphere overlaps the disk, as
is the case, when a non-Keplerian flow is present above the disk,  the
CI illumination parameter $f$ is not correlated with
color temperature and relatively high illumination is expected. 
In this section we will analyze the effect of
the illumination of the BMC region and derive certain constraints
on the structure of the accretion depending on the absolute value of
$f-$parameter  and its dependence on the color  temperature.

To analyze the illumination effect  we assume that the disk emits
isotropically and extends from the inner radius $r_{in}$ to the outer
radius $r_{out}$, while the outer radius of the CI atmosphere is
$r_{sp}$. Furthermore, we assume that $r_{sp}\leq r_{in}$.
In fact, the disk flux intercepted by the CI atmosphere with radius
$r_{sp}$   is calculated through the integral
\begin{equation}
F_{int}=\pi\int_{r_{in}}^{r_{out}}rdr\int_{-\pi/2}^{\pi/2}\cos{\varphi}
d\varphi \int^1_{\mu_{\ast}}(1-\mu^2)^{1/2}d\mu
\end{equation}
where $\mu_{\ast}=\sqrt{1-(r_{sp}/r)^2}$.  In the case where the disk
illuminates the slab   (with the half-width $\Delta r=r_{in}=3R_s$ and
the half-height $H$) situated in the center
~~$\mu_{\ast}=(r-r_{in})/\sqrt{(r-r_{in})^2+H^2}$.
Two internal integrals of Eq. (15) are calculated analytically.
The first one is equal to 2 and the second one is~~
$\theta_{\ast}-0.5\sin{2\theta_{\ast}}$,
where $\theta_{\ast}=\arccos\mu_{\ast}$.
The triple integral of Eq. (15) can also be estimated analytically
with accuracy better than 10\% for the spherical case with the
aforementioned assumption that $r_{sp}\leq r_{in}$:
\begin{equation}
F_{int}= {{2\pi}\over3} r_{sp}^3(r_{in}^{-1}-r_{out}^{-1}).
\end{equation}
In the case, when the part of the disk is overlapped  by the cloud,
  $F_{int}$ is calculated by integration from $r_{sp}$ to
$r_{out}$ followed by adding $\pi^2(r_{sp}^2-r_{in}^2)$,  the disk flux
emitted inside the atmosphere.
The fraction of the disk photons illuminating the CI  atmosphere is
\begin{equation}
f_{ill}= {{F_{int}p_{sc}}\over{\pi^2(r^2_{out}-r^2_{in})}}.
\end{equation}
where $p_{sc}$ is  a probability of scattering of the disk photons  in
the CI atmosphere of optical depth $\tau$, and $p_{sc} \sim
1-e^{-\tau}$.
The optical depth of the CI region is $\sim$ 1:
$\tau\sim \dot m[1.5^{-1/2}-3^{-1/2}]\approx 1$.    Here for the
optical depth  estimate we use formula (2) from TMK97  with an
assumption that $r_{out}=3R_s$ and $r_{in}=1.5R_s$ (TZ98).
But a smaller number of photons emerges after multiple scatterings
to form the hard tail of the spectrum (see Fig. 1).  We remind the
reader (see ST80, Titarchuk 1994b, and TMK97 for details ) 
that photons undergoing
multiple scatterings produce a specific space distribution
(in accordance to the first space  eigenfunction, see TZ98, Fig. 3) and
that fraction of  photons undergoing  multiple scattering in the plasma
cloud, $f_{ms}$ is related to the expansion coefficient of the space
source distribution over  the first space eigenfunction. In the simplest
case,  the uniform  source distribution $f_{ms}$ is approximately  0.8.

The parameter $f$,  used for the spectral fitting, can be estimated as:
\begin{equation}
f\sim {{f_{ill} f_{ms}A}\over{(1-f_{ms})f_{ill}A+\mu(1-f_{ill})}}
\end{equation}
where $f_{ill}$ is an illumination factor, $\mu$ is the cosine of
inclination angle of the disk and $A$ is a spherical albedo of
the CI atmosphere. The quantity $A$  can be estimated as
follows (e.g. Sobolev 1975):
\begin{equation}
A=1-{1\over{3\tau/4+1}}.
\end{equation}
This formula is obtained in the Eddington approximation
with an assumption of isotropic scattering at any event and
the pure absorptive inner boundary. Despite the Eddington
approximation the albedo formula (19) provides an accuracy $\sim$ 10\%
for any optical depth. For optical depth 1,  $A=0.42$, which  is
slightly higher than  value obtained by LT98 in the Monte Carlo simulations,
 $A_{MC}= 0.36$, because of  photon bending effects.

Now we present the numerical estimates of  $f-$parameter for
different cases:
The first one relates to the Monte Carlo simulations (LT98)
when $p_{sc}\sim 1-e^{-1}=0.63$, $f_{ms}=0.8$, $A=0.42$ and we assume
that $\mu\sim 1$. This gives $f\sim 0.5$ which is very close to
what we get from the application of our model to Monte Carlo simulated
spectra. It is worth noting here that in the limit of the high optical
depth (or mass accretion rate) $f\sim f_{ms}/(1-f_{ms})$ because
$p_{sc}$, $f_{ill}$, and $A$ converge to one.
Thus, in this particular case, the $f-$parameter  can approach or even
exceed unity.

\centerline{Editor: Place Figure 8 here}

In Figure 8 we present the observational and theoretical determinations
of the illumination parameter, $f$ versus the color temperature,
$T_{col}$ for two sources, GRO J1655-40 (solid circles) and
GRS 1739-278 (hollow circles). The theoretical determinations
the CI illumination have been made applying cylinder geometries, with
the half-height $H=3r_s$ (solid line),  and  spherical geometries
with $r_{sp}=3r_s$ (dotted line).
In both cases we assume that the inclination angle cosine $\mu=0.5$.
The dashed line describes the $f$-dependence on the color temperature,
when the analytical approximation of the illumination factor $f_{ill}$
(Eqs. 15-16) is used.

In the case of GRO J1655-40 all observational points on Fig.8 lie
in the region of $f-T_{col}$ space associated with a CI region
situated within a few Schwarzschild radii ($\sim 3~ r_s$).
But in the source GRS 1739-278 the converging inflow either shrink
to smaller radius ($f-$parameter is about 2\%),  or it overlaps the
underlying disk ($f-$parameter is $\gtorder 15\%$).
It is worth noting that for GRS 1915+105 the $f-
$parameter is $\gtorder 60\%$ and we can speculate that the
CI region covers a significant part of the disk. In fact, the
Comptonizing region must overlap the
inner region of the disk, in order to get such a high fraction of
the  seed photons from the disk.
But on the other hand, high values of  $f-$parameter
($\gtorder 10\%$) can be obtained, if the innermost  region of the disk
is puffed up to heights of  $\sim 3 R_s$ and the disk is seen nearly edge
on.

\section{Discussions and Conclusions}

Our analysis of broad-band X-ray spectral data for several soft X-ray
transients, representing a large fraction of the publicly available RXTE
archive for this type of sources, demonstrate that the bulk-motion
theory, even in its simplified analytical form, satisfactorily
approximates all of the data.

We would like to point out that in many cases,
consideration of  the BMC upscattering effects leads to lower
effective disk color temperatures in comparison with commonly used
models involving the sum of a black-body (or multi-color disk) plus
power-law component. Interpretation of our results does
not require  consideration  of black hole rotation, which has been
invoked earlier to explain reported  high temperatures (2-4 keV)
observed in accretion disks (Zhang et al. 1997a, Zhang, Cui \& Chen
1997b).

The bulk motion theory allows one to explain, in a self-consistent
manner, the origin of two-component spectra in high state on the basis
of simple geometrical assumptions and basic
physical principals. Study of parameter variations show the overall
correlation between the temperature of disk component and X-ray flux and
demonstrates the decrease of effective disk radius when the system
starts its  transmission to low spectral state.

The variation of the power law index could be induced either by
variations in mass accretion rate reflected by temperature changes (see
the  difference between spectra of GRS1739-278 obtained in March and in
May, 1996), or by a difference in mass accretion rate for the non-Keplerian
(advection) flow only (with the disk accretion remaining stable). The latter
interpretation, if confirmed would be a tool to study inflow/outflow
processes in the disk.  Radio data of good quality could support this
interpretation, if radio micro-bursts are found to be in coincidence
with the increase in the hard power law index. This is apparently the
case for GRS 1915+105 source (Mirabel et al. 1998).

For any reasonable hydrodynamical model the final stage of the
accretion, i.e. when any disk structure no longer exists, occurs in a
nearly free fall manner. The radiation pressure forces cannot stop
matter from falling in, because they are significantly below of the
Eddington limit. In the soft spectral state of the black hole systems
the whole system is cooled down to  temperatures of 10 keV or less.
In this case the thermal motion is not able to generate the hard
photons, which are seen in the observations of the soft  state. But for
any model of accretion flow -- advection dominated, as  considered by
(e.g.) Narayan \& Yi (1994) or Chakrabarti \& Titarchuk (1995),  or
optically thick disk, analyzed by Shakura \& Sunyaev (1973) -- the
accreting matter enters into free fall upon its final black hole.  It is
worth noting here that effects associated with the innermost
boundary were not treated in deriving the self-similar advection
dominated solution (Narayan \& Yi 1994).  It was not within the
scope of their work, the goal of which was to demonstrate the
existence of the hot solution for the  optically thin case, which may
be associated with the quiescent of the X-ray novae. If one  applies
the same technique of the self-similarity for the  optically thick case,
i.e. when the mass accretion rate is of order of  Eddington, one
derives the cold solution  (e.g.. Wang \& Zhou 1999). We assert that
in the context of this solution, the only possible mechanism for
generating the hard photons is the bulk-motion effect in the region
within a few gravitational radii (Titarchuk \& Zannias 1998, Laurent
\& Titarchuk 1998).  In fact, Esin, McClintock \& Narayan (1997)
note that the bulk motion effect could be very important for the hard
spectrum formation  in the soft-high state of the BH systems.

We wish to emphasize that the BMC theory does not preclude
other mechanisms contributing to spectral formation within the inner
accretion flow. It is possible that BMC model, alone can not reproduce
the power law spectrum beyond $\sim500$ keV, which
 may have been seen from GRO J1655-40 with OSSE (Grove et al. 1998;
Tomsick et al 1998). As we have noted however, we have examined these
data independently and it is our view claims of a  {\it detection}, as
opposed to an extension of a {\it fit} to the data, is not irrefutable.

Quite distinct from the high-soft state, the low-hard spectral state
does not appear to uniquely associated with black hole binaries.
Neutron star binaries, when observed in low-luminosity states, in
fact exhibit spectra which are strikingly similar to the black-
hole low-hard state (Barret \& Vedrenne 1994; Barret \& Grindlay 1995;
Barret et al. 1999). The best demonstrations of this come from
observations of X-ray bursters during periods of low-intensity. The high-energy
spectra are very well represented by thermal Comptonization models
as is the case for the low-hard state black holes. Non-bursting, high-
luminosity low-mass X-ray binaries containing weakly magnetized
neutron stars, on the other hand  they  show no indication of
sustaining high-energy emission. This seems to be a clear indication
that the high-soft state, and in particular the high-energy power-law
associated with that state, is directly tied to the physics of the black
hole environment. This, further, strongly suggestive that the
phenomena which underlie it are associated with the innermost
probable regions of the black hole system.

It is important to point out that bulk-motion spectral component can  be
presented in {\it the high state spectra of black hole systems},
because its existence is dependent on the presence of a black-hole
horizon - this is a natural ``drain'' allowing removal of extra energy
from the system.  In this sense the bulk-motion flow is an advection-dominated
flow, because the most of energy  is advected onto the black hole and
only a small percentage contributes to an increase in the number of emergent
X-ray photons.

One of the principal predictions of the theory -- that
two-component spectra are typical for BH binaries -- is an attribute
of  black hole physics only, and is not applicable to neutron star
binaries. This prediction seems to be in agreement with observational
data thus far obtained.

Some additional assumptions are necessary to extract the basic physical
parameters of the system from its X-ray spectra.  GRO J1655-40,  a
well-measured system, can be used as  a template to develop a formalism
allowing one to make estimations of black hole mass and a distance for
other sources based on their spectral properties only. For example,
using our derived distance-to-mass ratio $0.67 \pm 0.06$  and the distance 
determination presented by Mirabel \& Rodrigues
(1994)  $12.5\pm1.5$ kpc for GRS 1915+105 we  find that the BH mass in
this source is $18.6\pm 2.2$ solar masses.

We have shown that at least for GRO J1655-40 and GRS 1739-278,
that an extraction of the best-fit parameters in terms of BMC model
allows us to come to reach certain conclusions regarding the structure
of the accretion disk. Our results are in favor to the standard disk
model (SS73) modified to the electron scattering as an origin
of soft component of BH soft state spectra.

\vspace{0.2in}
\centerline{\bf{ ACKNOWLEDGMENTS}}
This work made use of services provided by the High-Energy Astrophysics
Science Archive Research Center (HEASARC) and the Compton Gamma Ray
Observatory Science Support Center at the NASA Goddard Space
Flight Center. L.T. acknowledges NASA for support under grant NAG5-7317
and thanks Jean Swank for encouragement in all aspects of this work.
The work of IKI co-authors have been supported partly
by RFBR grant 96-15-96343. Authors are very grateful to the referee for
the his/her interesting comments on the manuscript.

\newpage
\par
\noindent

\begin{figure}
\plotone{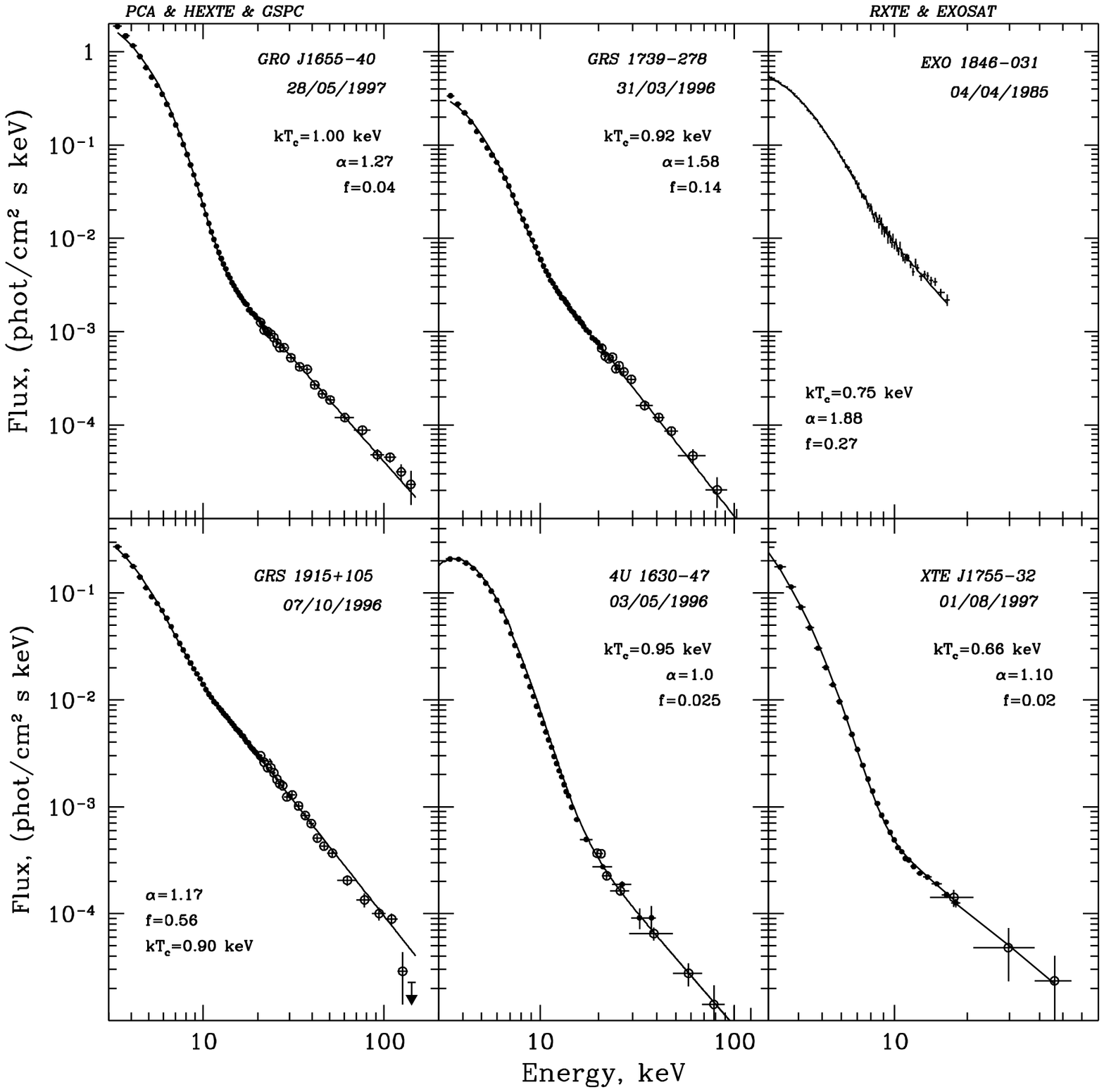}
\caption{Broad-band high-energy spectra of Galactic black hole
transient sources obtained by PCA and HEXTE experiments aboard
RXTE satellite (EXO1846-031 spectrum plotted on right top panel was
obtained by EXOSAT satellite). The axes are photon flux
(ph/cm$^2$/s/keV)
versus energy (keV). PCA data are marked by {\em solid circles},
HEXTE data - by {\em hollow circles}. Solid lines represent bulk-motion
model fits of the data. Source names, dates of observations and
related best-fit model parameters are written on the panels.
For XTE J1755-32 (right bottom panel) the upper limit for
the flux detected with HEXTE in high energy
band is shown by inclined line with arrow.}
\end{figure}

\begin{figure}
\plotone{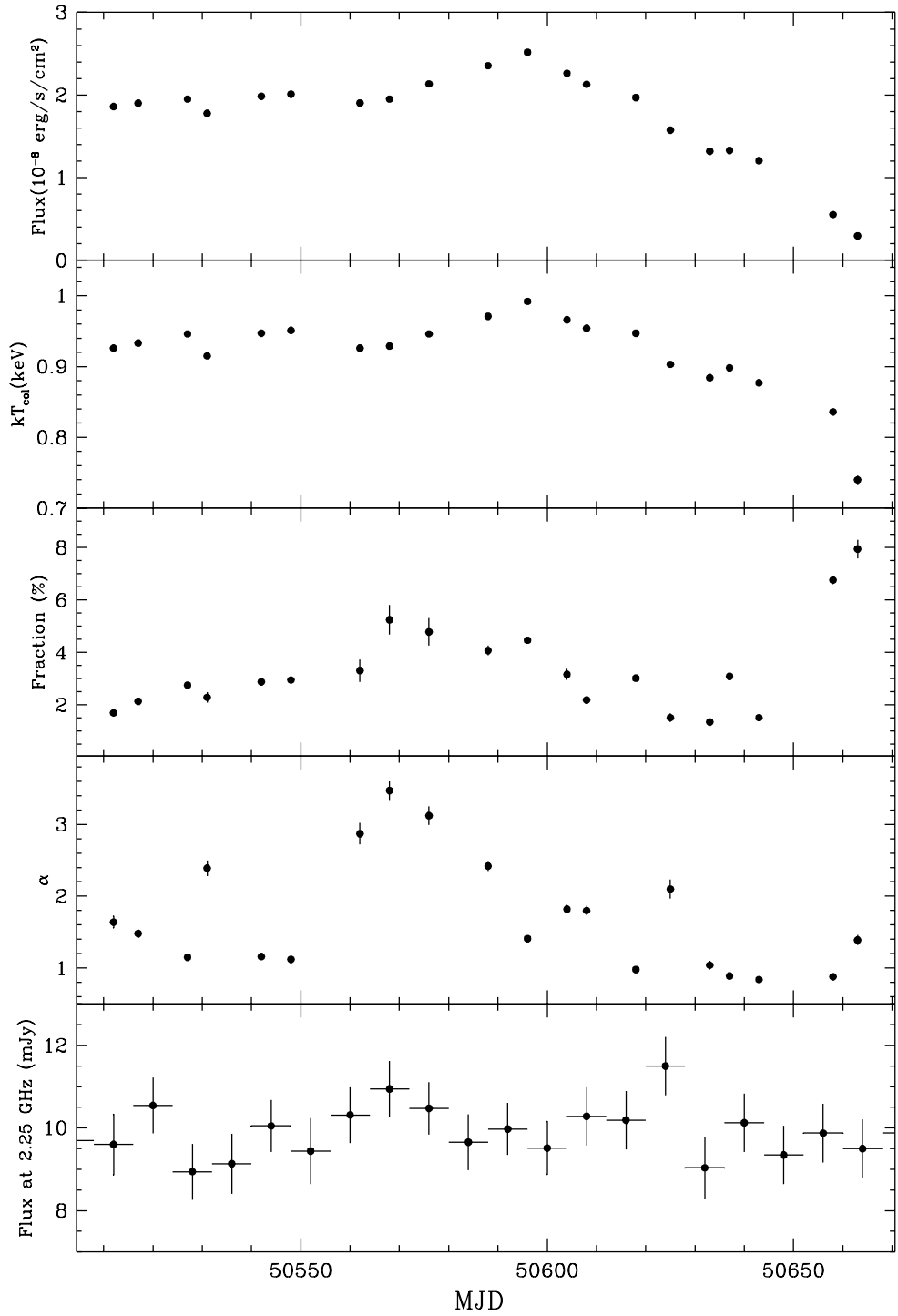}
\caption{Time variability plots for GRO J1655-40.  The following
data are plotted (from top to bottom): 1) Energy flux in 3-25 keV band
(PCA/RXTE data); 2-4) parameters of bulk motion model approximation (2 -
color temperature, 3 - fraction, 4 - power law index); 5) radio flux
density at 2.25 GHz (GBI data).
}

\end{figure}

\begin{figure}
\plottwo{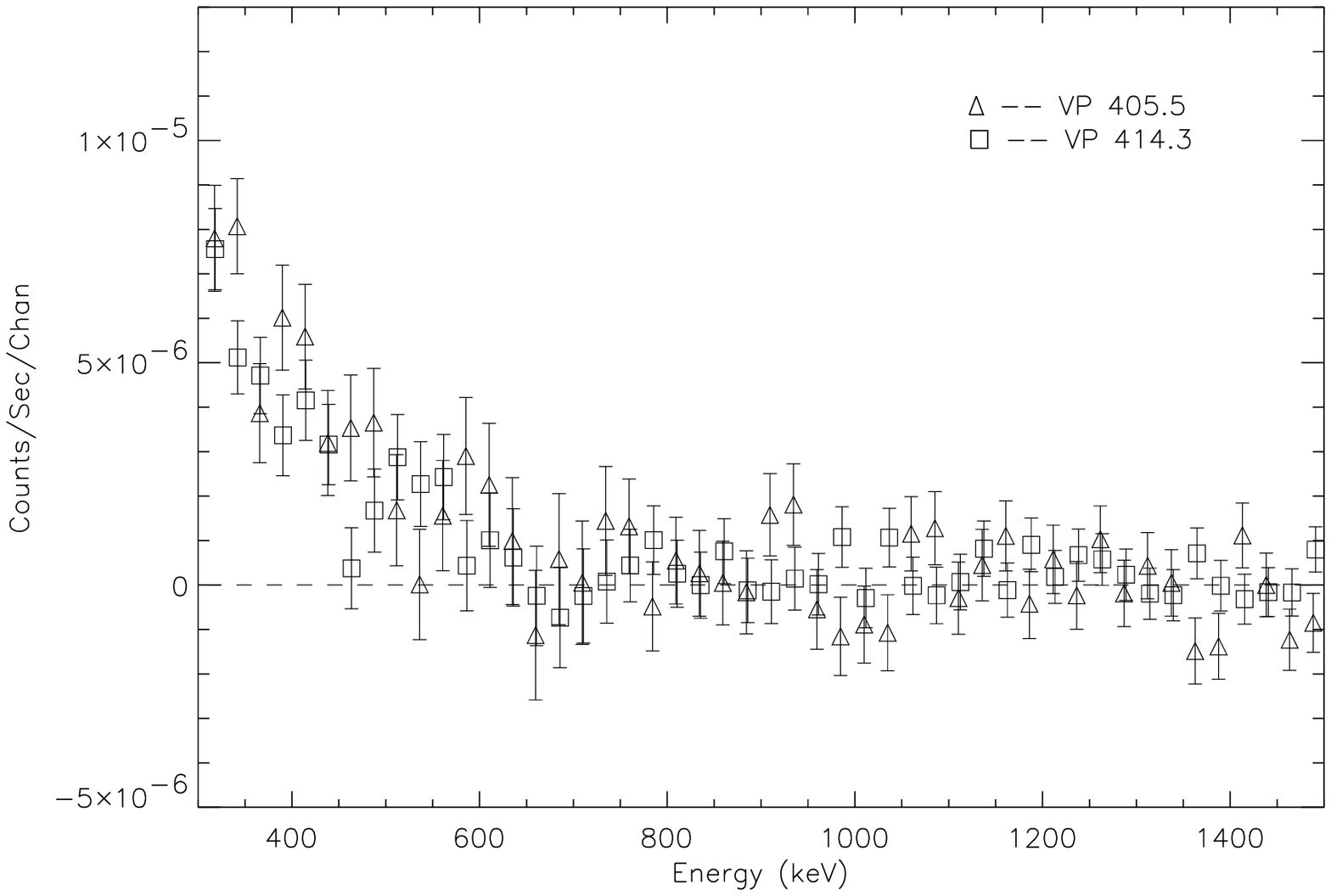}{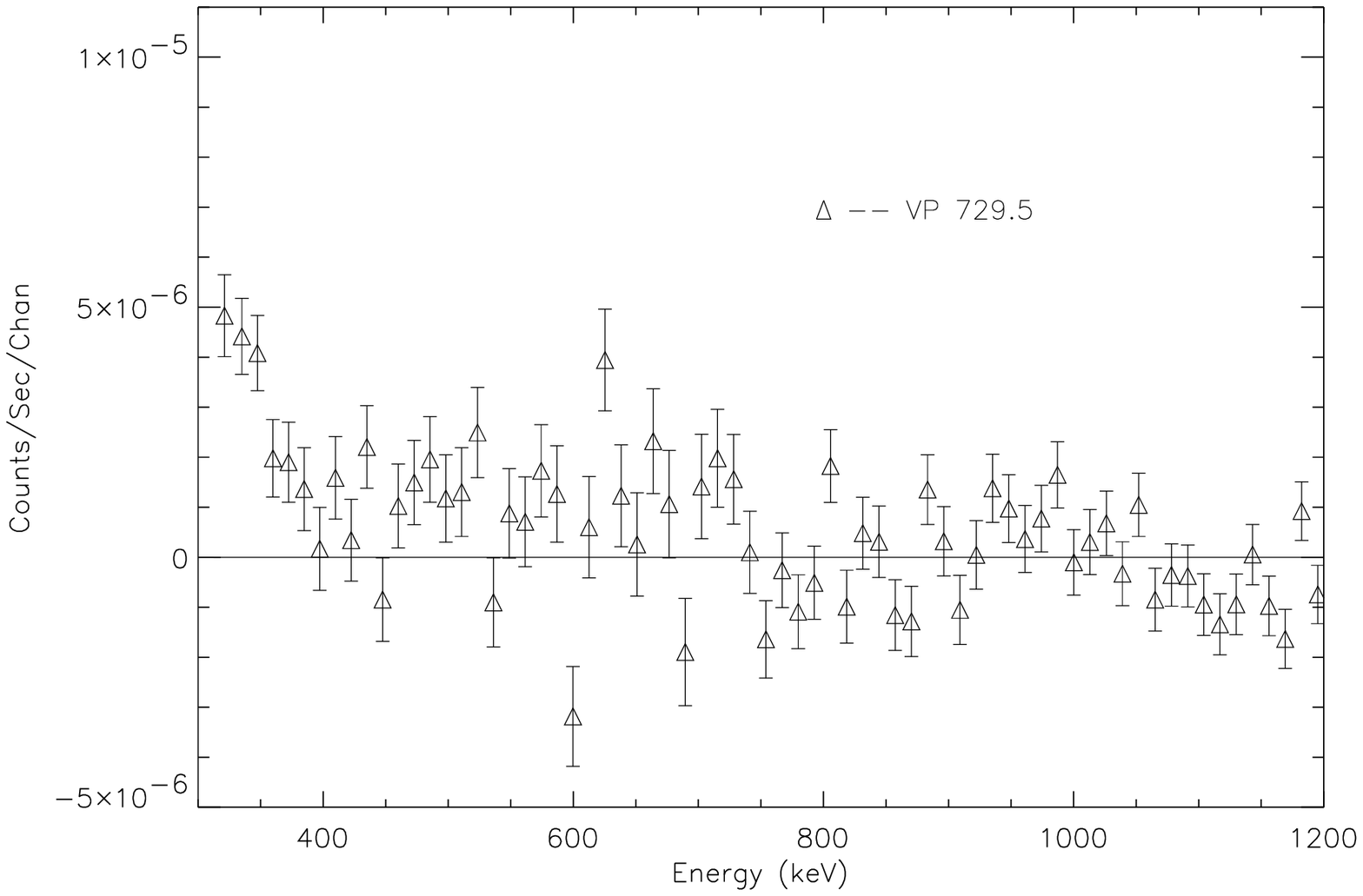}
\caption{(a) Summed detector count rates for channels
corresponding to energies above about 300~keV for 2 separate OSSE
observations of GRO~J1655-40 overlaying a zero-intensity model.
The dearth of significant counts above $\sim500$~keV are consistent with
BMC model predictions. (b) a similar plot for the recently discovered
XTE~J1550-564. These two sources are among the brightest examples
of high-soft-state spectra above 100~keV.
}
\end{figure}

\begin{figure}
\plotone{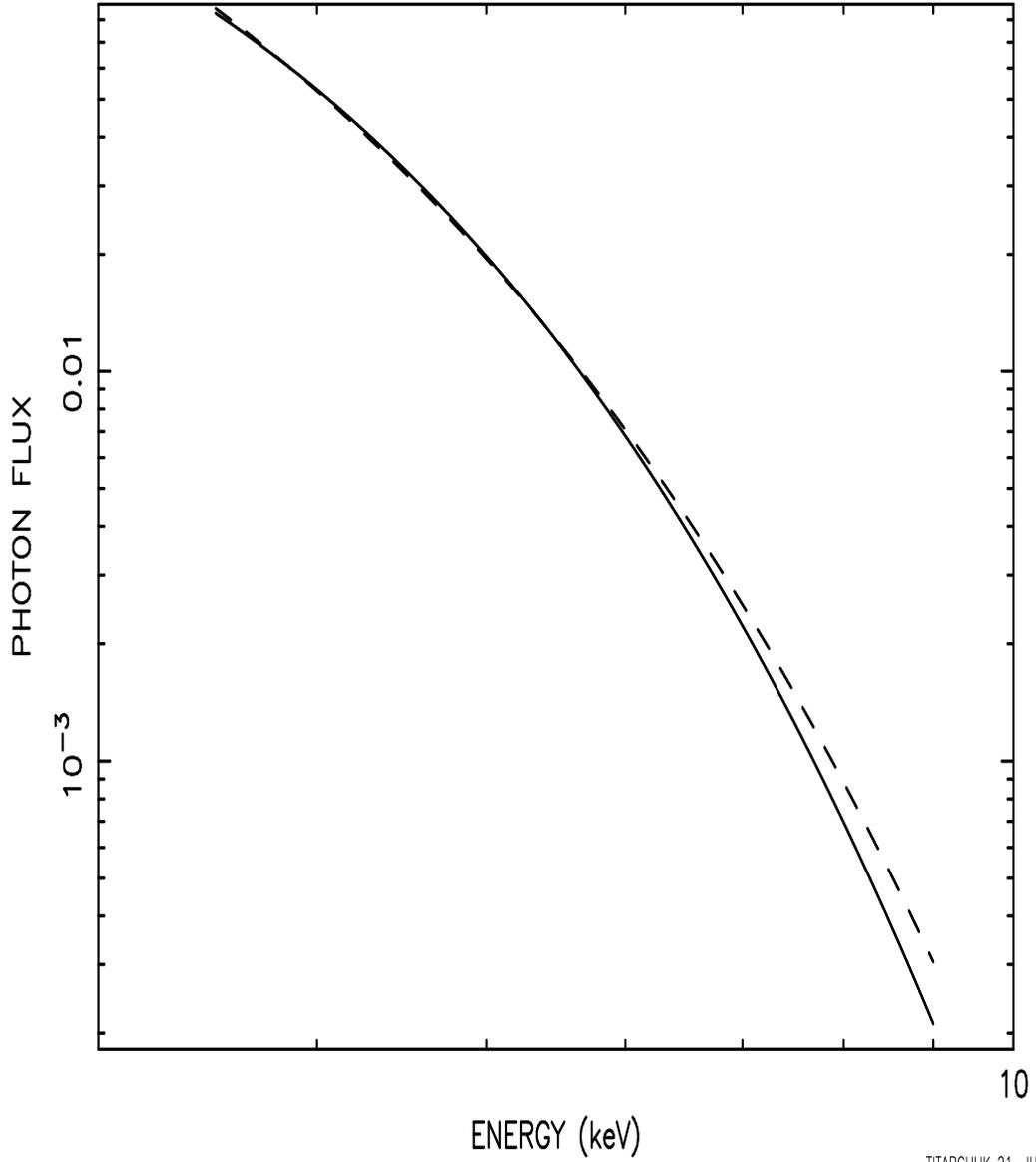}
\caption{ Comparison of  multicolor blackbody disk spectrum with
maximal disk temperature $T_{max}=0.809$ keV (dashed line)
and the closest single blackbody
spectrum (in terms of the least square method)
with the color  temperature $T_{col}=0.7$ keV (solid line).
}
\end{figure}

\begin{figure}
\plotone{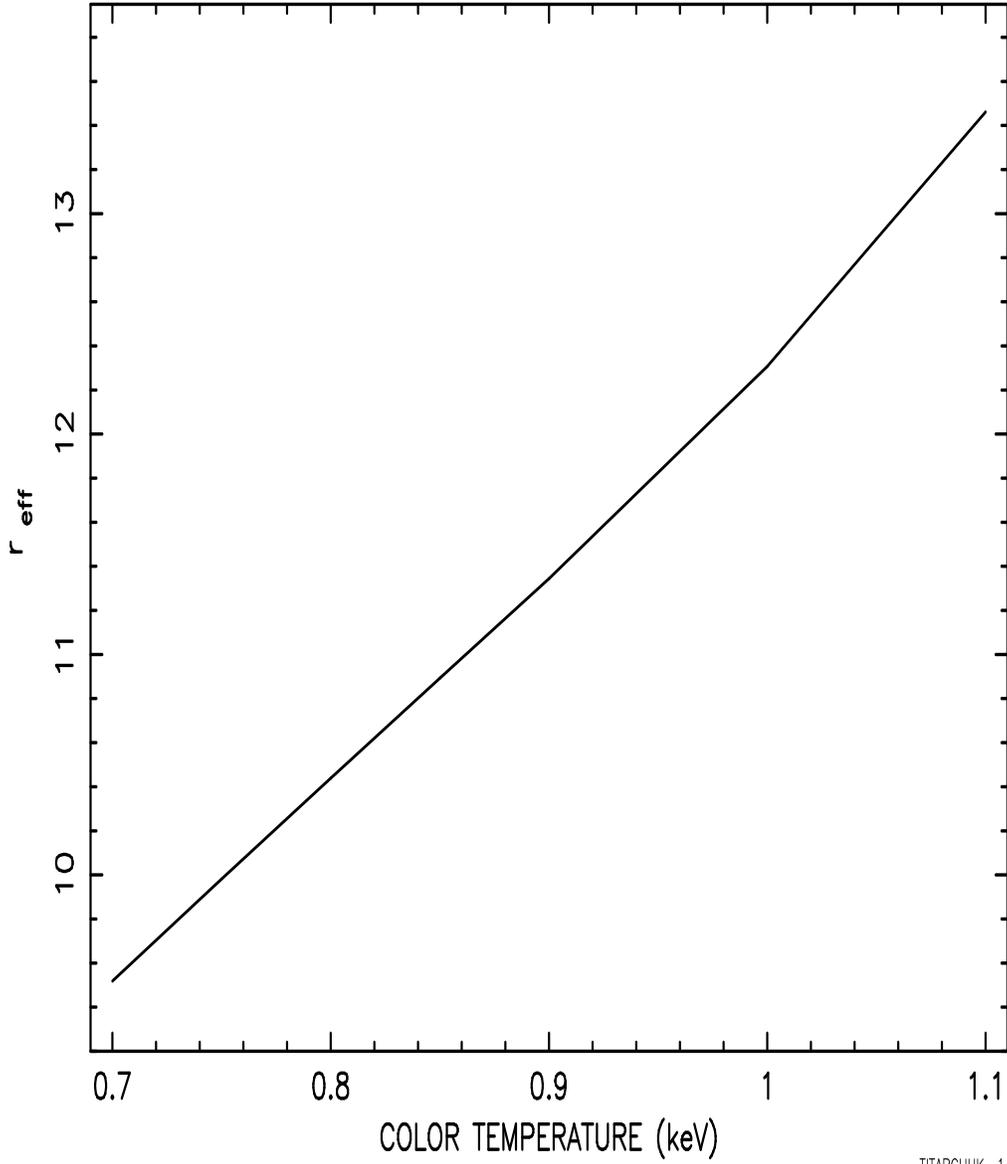}
\caption{Plot of  the theoretical dependence of the
 effective radius of the disk emission region
$r_{eff}$ vs color temperature $T_{col}$ (keV).
}
\end{figure}

\begin{figure}
\plotone{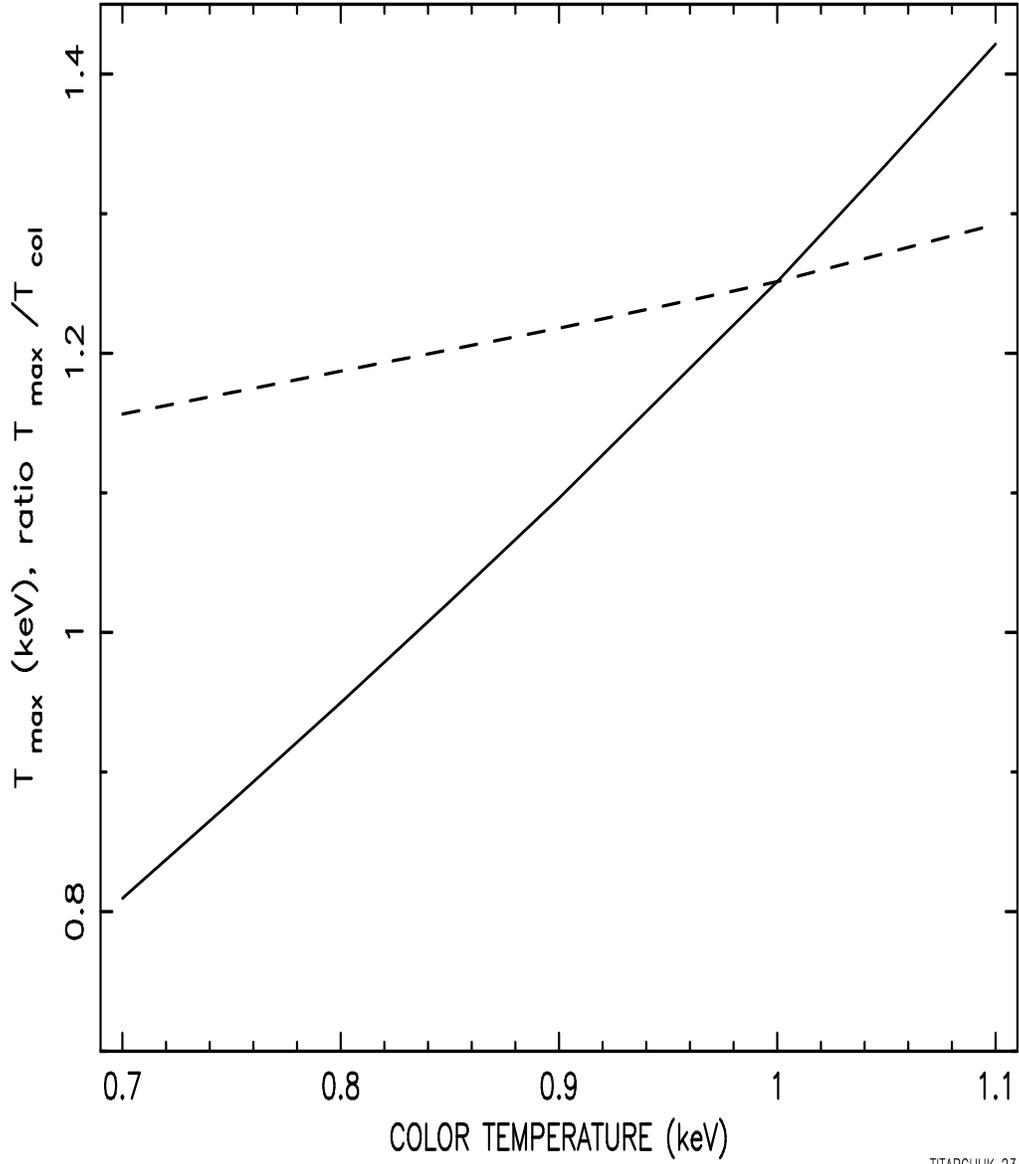}
\caption{Plot of  the theoretical dependence of the maximal
disk temperature (solid line) and ratio $T_{max}/T_{col}$ (dashed line)
vs  color temperature $T_{col}$ (keV).
}
\end{figure}

\begin{figure}
\plotone{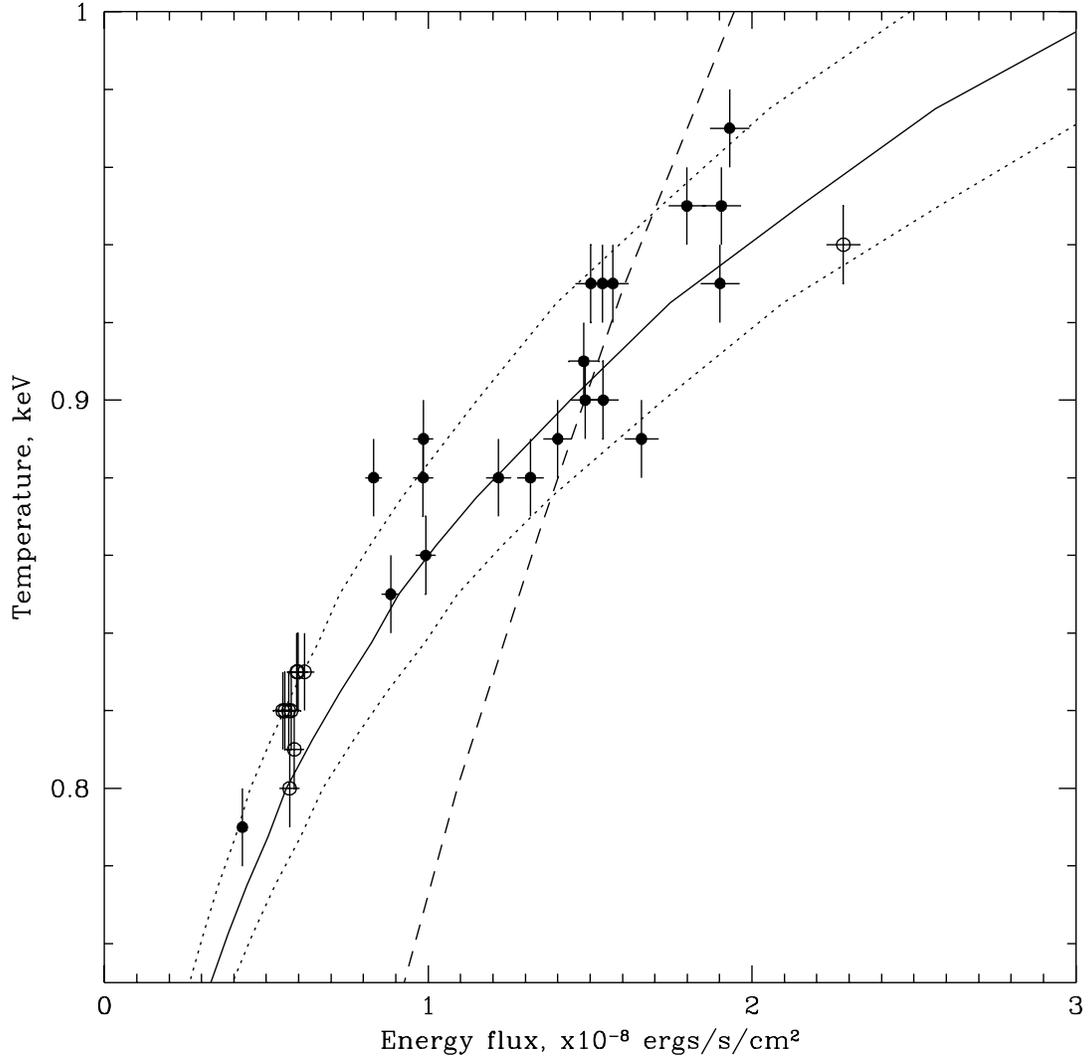}
\caption{ Plot of  color temperature $T_{col}$ (keV) vs energy
flux dependance plot. Solid points
are bulk motion fit parameters for GRO J1655-40, hollow circles - for
GRS 1739-278. Solid curve is the theoretical dependence for bulk motion
model, dashed curve - for black-body model.  Absolute normalizations of
flux
for theoretical curves and GRS1739-278 data are chosen to agree better
with GRO J1655-40 data. Dotted curves are having normalization 0.8 and
1.2 relative to solid (best fit) curve.
}
\end{figure}

\begin{figure}
\plotone{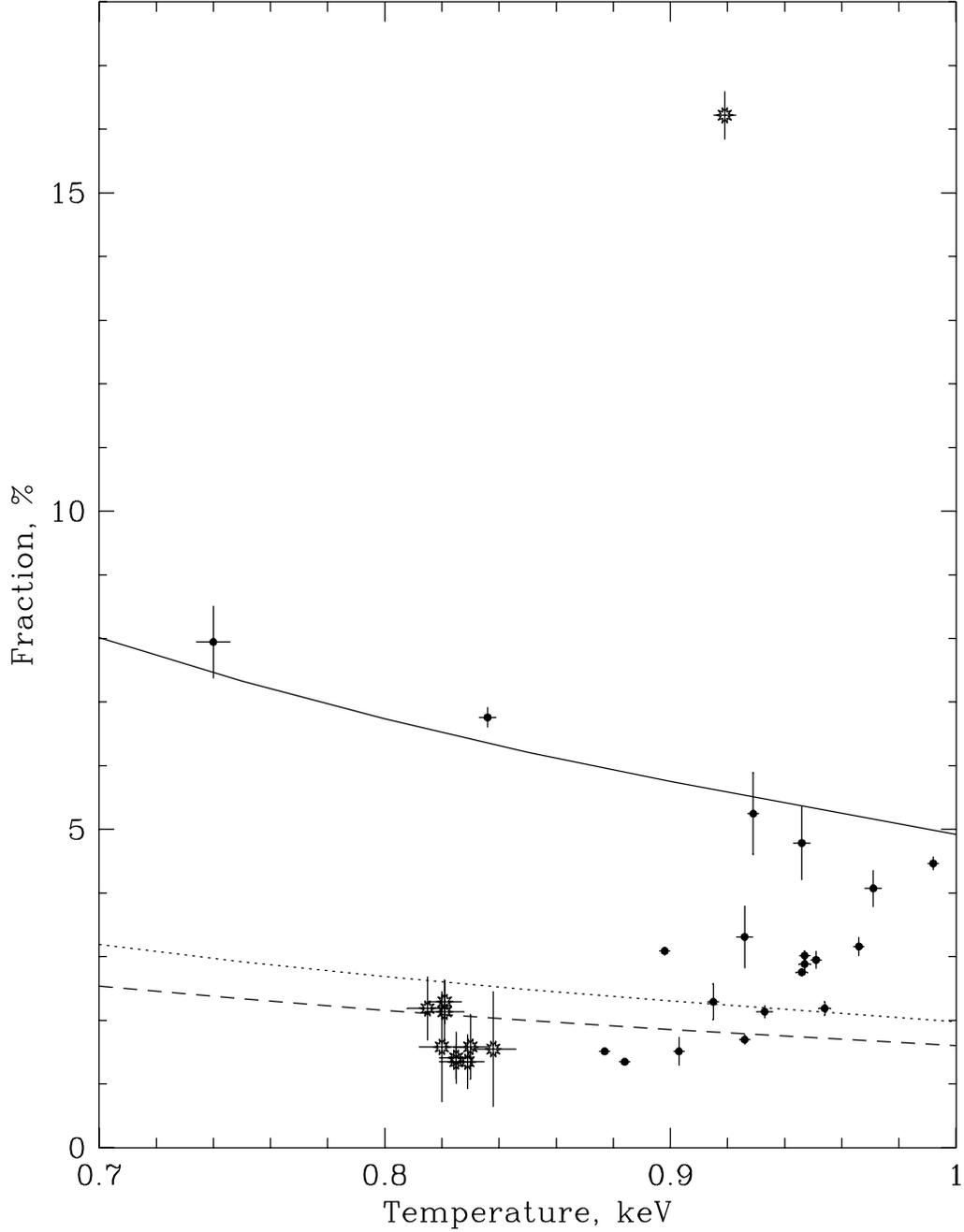}
\caption{Plot of  illumination fraction parameter $f$ vs color
 temperature $T_{col}$. Solid points are
bulk motion fit parameters for GRO J1655-40, hollow circles - for GRS
1739-278. Solid curve is the theoretical dependence for illumination of
the central region in the case of cylindrical geometry, dotted line
is the same  in the case of  spherical geometry and dashed curve is
the analytical approximation for spherical geometry (see text).
}
\end{figure}

\newpage

\begin{deluxetable}{ccccc}
\small
\tablecolumns{5}
\tablecaption{}
\tablehead{
\colhead{$\#$}&
\colhead{Date obs.}&
\colhead{Time start, UT}&
\colhead{Time end, UT}&
\colhead{PCA$^a$/HEXTE$^b$ live time, s}\\
\cline{1-5}\\
\multicolumn{5}{c}{\it XTE J1755-324}
}
\startdata
1&29/07/97&05:17:20&06:54:24&768\\
2&01/08/97&05:19:12&06:58:24&1993\\
\cutinhead{\it GRS 1739-278}
1&31/03/96&18:08:48&20:50:56&5215/2066\\
2&10/05/96&07:08:00&07:56:00&1365\\
3&11/05/96&18:58:56&20:05:52&2421\\
4&12/05/96&03:23:28&03:45:52&1117\\
5&13/05/96&17:05:20&17:50:56&2110\\
6&14/05/96&13:43:28&14:48:00&2886\\
7&15/05/96&17:04:32&18:25:52&325\\
8&16/05/96&23:48:32&24:34:56&1691\\
9&17/05/96&12:24:32&13:26:56&3398\\
10&29/05/96&02:47:28&03:57:52&2529\\
\cutinhead{\it EXO 1846-031}
1&04/04/85&00:06:56&02:32:32&6959$^c$\\
\enddata
\tablenotetext{a}{Deadtime corrected value}
\tablenotetext{b}{For each cluster of HEXTE}
\tablenotetext{c}{GSPC time}
\end{deluxetable}
















\begin{deluxetable}{ccccccc}
\small
\tablecolumns{7}
\tablecaption{}
\tablehead{
\colhead{Date} &
\colhead{$T_{col}$, keV} &
\colhead{$\alpha$}&
\colhead{$f, \%$} &
\colhead{$A_N$}&
\colhead{$F_{5-25 keV}^a$}&
\colhead{$\chi^2$/dof}\\
\cline{1-7}\\
\multicolumn{7}{c}{\it GRS 1739-278}
}
\startdata
31/03/96&$ 0.94 \pm 0.01$&$1.39\pm0.04$&$13.7\pm0.5$&$1.33 \pm0.1$&$3.64

\pm0.07$&22.8/36\nl
10/05/96&$ 0.83 \pm 0.01$&$1.07\pm0.25$&$1.15\pm0.15$&$1.41
\pm0.15$&$1.10\pm0.02$&26.5/36\nl
11/05/96&$ 0.83 \pm 0.01$&$0.70\pm0.19$&$0.89\pm0.07$&$1.41
\pm0.15$&$1.06 \pm0.02$&34.1/36\nl
12/05/96&$ 0.83 \pm 0.01$&$0.94\pm0.31$&$1.05\pm0.23$&$1.43
\pm0.03$&$1.05 \pm0.02$&27.5/36\nl
13/05/96&$ 0.82 \pm 0.01$&$1.13\pm0.23$&$1.16\pm0.21$&$1.44
\pm0.04$&$1.00 \pm0.02$&28.3/36\nl
14/05/96&$ 0.82 \pm 0.01$&$1.28\pm0.17$&$1.57\pm0.22$&$1.46
\pm0.04$&$1.02 \pm0.02$&30.3/36\nl
15/05/96&$ 0.80 \pm 0.01$&$1.37\pm0.30$&$1.53\pm0.39$&$1.66
\pm0.06$&$1.00 \pm0.02$&19.6/36\nl
17/05/96&$ 0.82 \pm 0.01$&$0.77\pm0.21$&$1.15\pm0.12$&$1.41
\pm0.04$&$1.00 \pm0.02$&25.6/36\nl
17/05/96&$ 0.82 \pm 0.01$&$0.82\pm0.14$&$1.21\pm0.10$&$1.41
\pm0.04$&$0.98 \pm0.01$&28.3/36\nl
29/05/96&$ 0.81 \pm 0.01$&$1.25\pm0.12$&$1.99\pm0.19$&$1.65
\pm0.04$&$1.05 \pm0.02$&30.5/36\nl
\cutinhead{\it XTE J1755--324}
29/07/97&$ 0.71 \pm 0.01$&$1.02\pm0.14$&$2.42\pm0.23$&$1.18
\pm0.06$&$0.29 \pm0.01$&29.8/36\nl
1/08/97&$ 0.72 \pm 0.01$&$0.89\pm0.08$&$2.82\pm0.15$&$0.933
\pm0.03$&$0.30 \pm0.01$&32.5/36\nl
\cutinhead{\it EXO 1846--031}
04/04/85&$ 0.75 \pm 0.01$&$1.88\pm0.11$&$27.\pm3.$&$3.51 \pm0.08$&
$9.15\pm0.09^b$ &237/204\nl
\enddata
\tablenotetext{a}{spectral flux in units of 10$^{-9}$ {\it ergs
s$^{-1}$ cm$^{-2}$}}
\tablenotetext{b}{hydrogen absorption column for this spectrum is
$N_{HL}=2.2\pm0.2 \times 10^{22} $ cm$^{-2}$ and the energy band is
2-20 keV}
\end{deluxetable}

\begin{deluxetable}{ccccccc}
\small
\tablecolumns{7}
\tablecaption{}
\tablehead{
\colhead{Date}&
\colhead{$T_{col}$, keV}&
\colhead{$\alpha$}&
\colhead{$f, \%$}&
\colhead{$A_N$}&
\colhead{$F_{5-20 keV}^a$}&
\colhead{$\chi^2_{36}$}\\
\cline{1-7}\\
\multicolumn{7}{c}{\it GRO J1655-40}
}
\startdata
05/01/97&$ 0.89 \pm 0.01$&$0.95\pm0.02$&$10.\pm0.1$&$3.13 \pm
0.03$&$6.14 \pm0.12$&29.9\nl
12/01/97&$ 0.88 \pm 0.01$&$0.91\pm0.02$&$5.3\pm0.01$&$3.21 \pm
0.02$&$4.65 \pm0.09$&37.9\nl
20/01/97&$ 0.79 \pm 0.01$&$0.94\pm0.07$&$0.9\pm0.01$&$3.69 \pm
0.04$&$1.80 \pm0.03$&31.6\nl
26/02/97&$ 0.88 \pm 0.01$&$2.61\pm0.09$&$2.6\pm0.2$&$5.36 \pm
0.04$&$6.27 \pm0.12$&37.5\nl
10/03/97&$ 0.91 \pm 0.01$&$1.5\pm0.07$&$2.1\pm0.1$&$4.86 \pm 0.04$&$7.61
\pm0.15$&31.4\nl
20/03/97&$ 0.93 \pm 0.01$&$1.21\pm0.03$&$2.9\pm0.01$&$4.59 \pm
0.03$&$8.27 \pm0.16$&34.5\nl
24/03/97&$ 0.89 \pm 0.01$&$2.59\pm0.11$&$2.7\pm0.2$&$5.29 \pm
0.05$&$6.82 \pm0.13$&31.4\nl
10/04/97&$ 0.93 \pm 0.01$&$1.16\pm0.04$&$3.2\pm0.1$&$4.5 \pm 0.04$&$8.58
\pm0.17$&24.4\nl
24/04/97&$ 0.89 \pm 0.01$&$3.84\pm0.21$&$7.0\pm1.2$&$5.18 \pm
0.05$&$7.44 \pm0.14$&28.0\nl
30/04/97&$ 0.90 \pm 0.01$&$3.68\pm0.15$&$6.5\pm0.8$&$5.4 \pm 0.04$&$7.78

\pm0.15$&23.1\nl
12/05/97&$ 0.89 \pm 0.01$&$5.36\pm0.14$&$26.3\pm3.5$&$5.1 \pm
0.06$&$8.88 \pm0.17$&20.1\nl
20/05/97&$ 0.94 \pm 0.01$&$2.78\pm0.08$&$5.6\pm0.4$&$4.81 \pm
0.04$&$10.2 \pm0.2$&15.6\nl
28/05/97&$ 0.97 \pm 0.01$&$1.55\pm0.04$&$4.9\pm0.1$&$4.26 \pm
0.04$&$11.0 \pm0.2$&30.6\nl
05/06/97&$ 0.94 \pm 0.01$&$1.98\pm0.06$&$3.4\pm0.2$&$4.75 \pm
0.04$&$9.63 \pm0.19$&17.6\nl
09/06/97&$ 0.93 \pm 0.01$&$2.54\pm0.09$&$3.9\pm0.3$&$4.82 \pm
0.04$&$8.61 \pm0.17$&17.2\nl
19/06/97&$ 0.93 \pm 0.01$&$1.01\pm0.04$&$3.0\pm0.1$&$4.36 \pm
0.04$&$8.16 \pm0.16$&17.2\nl
26/06/97&$ 0.88 \pm 0.01$&$1.89\pm0.20$&$1.2\pm0.2$&$5.02 \pm
0.06$&$5.84 \pm0.10$&28.1\nl
04/07/97&$ 0.86 \pm 0.01$&$1.07\pm0.06$&$1.3\pm0.01$&$4.66 \pm
0.04$&$4.73 \pm0.09$&38.6\nl
08/07/97&$ 0.88 \pm 0.01$&$0.92\pm0.02$&$3.4\pm0.01$&$3.93 \pm
0.04$&$5.17 \pm0.10$&23.6\nl
14/07/97&$ 0.86 \pm 0.01$&$0.92\pm0.05$&$1.5\pm0.01$&$4.33 \pm
0.04$&$4.23 \pm0.08$&37.2\nl

\enddata
\tablenotetext{a}{Spectral flux in units of 10$^{-9}$ {\it ergs
s$^{-1}$ cm$^{-2}$}}
\end{deluxetable}

\begin{deluxetable}{ccccccccccc}
\small
\tablecolumns{11}
\tablecaption{}

\tablehead{
\colhead{Source} &
\multicolumn{4}{c}{Spectral fit parameters} &
\multicolumn{6}{c}{Physical parameter estimations}\\
\cline{2-5}\cline{6-11}\\
&
\colhead{$A_N$}&
\colhead{$T_{col}$}&
\colhead{$f$}&
\colhead{$\alpha$}&
\colhead{$T_{max}$}&
\colhead{$r_{eff}$}&
\colhead{$\dot m_{disk}$}&
\colhead{$m^{\star}$}&
\colhead{$d^{\star}$}&
\colhead{$L_{37}^{\star}$}
}

\startdata
GRS 1915+105$^a$ & 3.0 & 0.9 & 0.72 & 1.7 & 1.1 & 11.5 &$\gtorder2$ &
$\gtorder9.7$ &$\gtorder6.5$ & $\gtorder5.1$\\
GRS 1739-278 & 1.33 & 0.94 & 0.14 & 1.4 & 1.16 & 11.8 &$\gtorder2$ &
$\gtorder8.6$ & $\gtorder7.7$ & $\gtorder5.$\\
XTE J1755-324 & 1.18 & 0.71 & 0.02 & 1.0 & 0.82 & 9.6 &$\sim0.5$ &
$\sim8.4$ & $\sim 8$ & $\sim1.0$\\
4U 1630-47 & 2.27 & 0.85 & 0.025 & 1.38 & 1.02 & 10.9 & $\gtorder2$ &
$\gtorder14.3$ & $\gtorder 10.3$ &$\gtorder7.9$\\
EXO 1846-031 & 3.5 & 0.75 & 0.27 & 1.88 & 0.88 & 9.9 &$\gtorder1$ &
$\gtorder 12$ & $\gtorder6$ & $\gtorder2.8$
\enddata

\tablenotetext{a}{Fit parameters for GRS1915+105 are taken from ShT98}
\tablenotetext{b}{Fit parameters for 4U 1630-47 were obtained using
``diluted'' black-body fit for soft component of the spectrum (see \S
4)}
\end{deluxetable}

\end{document}